\documentclass[12pt]{article}
\def\macrosPb{}
\usepackage{amsfonts}
\usepackage{amsmath,amssymb,amsthm}
\usepackage{appendix}
\usepackage{bbm} 
\usepackage{amsbsy}
\usepackage{enumerate}
\usepackage{cite}

\def\macrosH{}
\def\macrosHarxiv

\InputIfFileExists{./macros_local.tex}{}{}

\ifdefined\macrosPa
  \usepackage[textwidth=465pt,textheight=650pt,centering]{geometry} 
\else\ifdefined\macrosPb
  \usepackage[textwidth=500pt,textheight=650pt,centering]{geometry} 
\fi\fi

\ifdefined\macrosS


  \usepackage{mathptmx}
  \DeclareMathAlphabet{\mathcal}{OMS}{cmsy}{m}{n}
\fi

\ifdefined\macrosBirk
\else
\usepackage[dvips]{graphicx}
\fi

\ifdefined\macrosSB

\def\UseSection{
        \numberwithin{equation}{section}
	\theoremstyle{plain}
        \newtheorem{theorem}    {Theorem}[section]
        \DefineTheorems 
}

\def\DefineTheorems{
	
	\newtheorem{lemma}      [theorem] {Lemma}
	
	\newtheorem{prop}       [theorem] {Proposition}
	
	\newtheorem{cor}        [theorem] {Corollary}

	\theoremstyle{definition}
	\newtheorem{defn}       [theorem] {Definition}

	\theoremstyle{definition}

}

\newcommand{\bt}   {\begin{theorem}}
\newcommand{\et}   {\end  {theorem}}
\newcommand{\bl}   {\begin{lemma}}
\newcommand{\el}   {\end  {lemma}}
\newcommand{\bp}   {\begin{prop}}
\newcommand{\ep}   {\end  {prop}}
\newcommand{\bc}   {\begin{cor}}
\newcommand{\ec}   {\end  {cor}}
\newcommand{\bd}   {\begin{defn}}
\newcommand{\ed}   {\end  {defn}}

\newcommand{\ba}   {\begin{array}}
\newcommand{\ea}   {\end  {array}}
\newcommand{\be}   {\begin{enumerate}}
\newcommand{\ee}   {\end  {enumerate}}
\newcommand{\bi}   {\begin{itemize}}
\newcommand{\ei}   {\end  {itemize}}

\def\eq#1\en{\begin{equation}#1\end{equation}}  
\def\eqsplit#1\ensplit{
	\begin{equation}\begin{split}#1\end{split}\end{equation}
	}
\def\eqalign#1\enalign{
	\begin{align}#1\end{align}
	}
\def\eqmul#1\enmul{
	\begin{multline}#1\end{multline}
	}
\newcommand{\eqarrstar} {\begin{eqnarray*}} 
\newcommand{\enarrstar} {\end{eqnarray*}} 
\newcommand{\eqarray}   {\begin{eqnarray}} 
\newcommand{\enarray}   {\end{eqnarray}} 
 
\makeatletter
\newcommand{\labelcounter}[2]{{%
	\stepcounter{#1}
	\protected@write\@auxout{}%
	{\string\newlabel{#2}{{\csname the#1\endcsname}{\thepage}}}%
	{\ref{#2}}
	}}
\makeatother
\newcommand{\Rbold} {{\mathbb R}}

\newcommand{\Zbold} {{\mathbb Z}}

\newcommand{\Scal}   {\mathcal{S}}

\newcommand{\spose}[1] {{\hbox to 0pt{#1\hss}} }
\newcommand{\ltapprox} {\mathrel{\spose{\lower 3pt\hbox{$\mathchar"218$}}
 \raise 2.0pt\hbox{$\mathchar"13C$}}}
\newcommand{\gtapprox} {\mathrel{\spose{\lower 3pt\hbox{$\mathchar"218$}}
 \raise 2.0pt\hbox{$\mathchar"13E$}}}

\else
\fi

\UseSection   
\usepackage[usenames]{color}

\definecolor{bw}{RGB}{240, 120, 0}
\definecolor{at}{rgb}{0.0, 0.5, 0.0} 

\renewcommand{\to} {\rightarrow}

\newcommand{\R}{\Rbold}
\newcommand{\Z}{\Zbold}

\newcommand{\1}{\mathbbm{1}}

\newcommand{\Ex}{\mathbb{E}}

\ifdefined\macrosSB \else

\fi

\ifdefined\macrosH
  \usepackage{xr-hyper}
  \usepackage{hyperref}
  \hypersetup{hypertexnames=false}
  \hypersetup{colorlinks,citecolor=blue,linkcolor=blue}  

\else\ifdefined\macrosHarxiv
  \usepackage{xr-hyper}
  \usepackage{hyperref}
  \hypersetup{hypertexnames=false, hidelinks}

\else
  
  \usepackage{xr}
\fi\fi

\hypersetup{
    colorlinks,
    citecolor=black,
    filecolor=black,
    linkcolor=black,
    urlcolor=black
}

 \title {
   Self-avoiding walk, spin systems, and renormalization
 }

 \author{
   Gordon Slade\thanks{Department of Mathematics,
     University of British Columbia,
     Vancouver, BC, Canada V6T 1Z2.
     E-mail: {\tt slade@math.ubc.ca}.}}

\begin{document}
\maketitle

\begin{abstract}
The self-avoiding walk, and lattice spin systems such as the
$\varphi^4$ model, are models
of interest both in mathematics and in physics.
Many of their important mathematical problems remain unsolved,
particularly those involving critical exponents.
We survey some of these problems, and report on
recent advances in their mathematical understanding via a rigorous
nonperturbative renormalization group method.
\end{abstract}

\section{Introduction}

The self-avoiding walk (SAW) is a combinatorial
model of lattice paths without self-intersections.
In addition to its intrinsic mathematical interest, it arises in
polymer science as a model of linear polymers, and in statistical mechanics
as a model that exhibits critical behaviour.  The mathematical problems
associated to the SAW are notoriously difficult and there remain longstanding
unsolved problems that are central to the subject.  A closely related model
is the weakly self-avoiding walk (WSAW),
which is predicted to exhibit the same critical behaviour as
the SAW.

The critical behaviour of the SAW or WSAW is expressed in terms of
critical exponents, which have a qualitative and quantitative relationship
with the critical exponents in models of ferromagnetism including
the Ising and $|\varphi|^4$ spin models.
Within physics, the critical exponents are well understood,
but they nevertheless present deep mathematical problems.

This article is a review of recent mathematical results about critical
exponents for the WSAW and $|\varphi|^4$ models,
with focus on the critical
behaviour of the susceptibility.
Some background on the SAW
and Ising models is provided for motivation and context.
The results we present
involve a unified treatment of the WSAW and $|\varphi|^4$
models, via an exact relation between the WSAW and a ``zero-component''
$|\varphi|^4$ model.
The proofs are based on a rigorous version of Wilson's renormalization group
(RG) approach. We provide an introduction to the RG method from the
perspective of a mathematician.

\section{Self-avoiding walk}
\label{sec:SAW}

We discuss the SAW and WSAW models, as well as their long-range versions.
The emphasis is on the critical behaviour, particularly for the susceptibility.

\subsection{Strictly self-avoiding walk  (SAW)}
\label{sec:SSAW}

\subsubsection{Universality and scale invariance}

\begin{figure}[h]
\centering{
\includegraphics[scale=0.13]{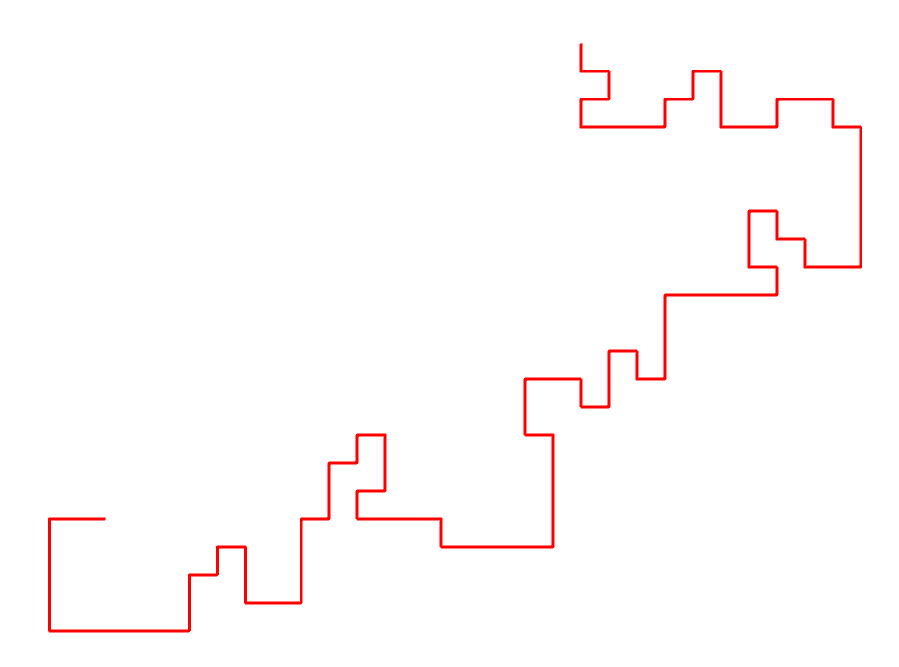}
\hspace{18mm}
\includegraphics[scale = 0.013]{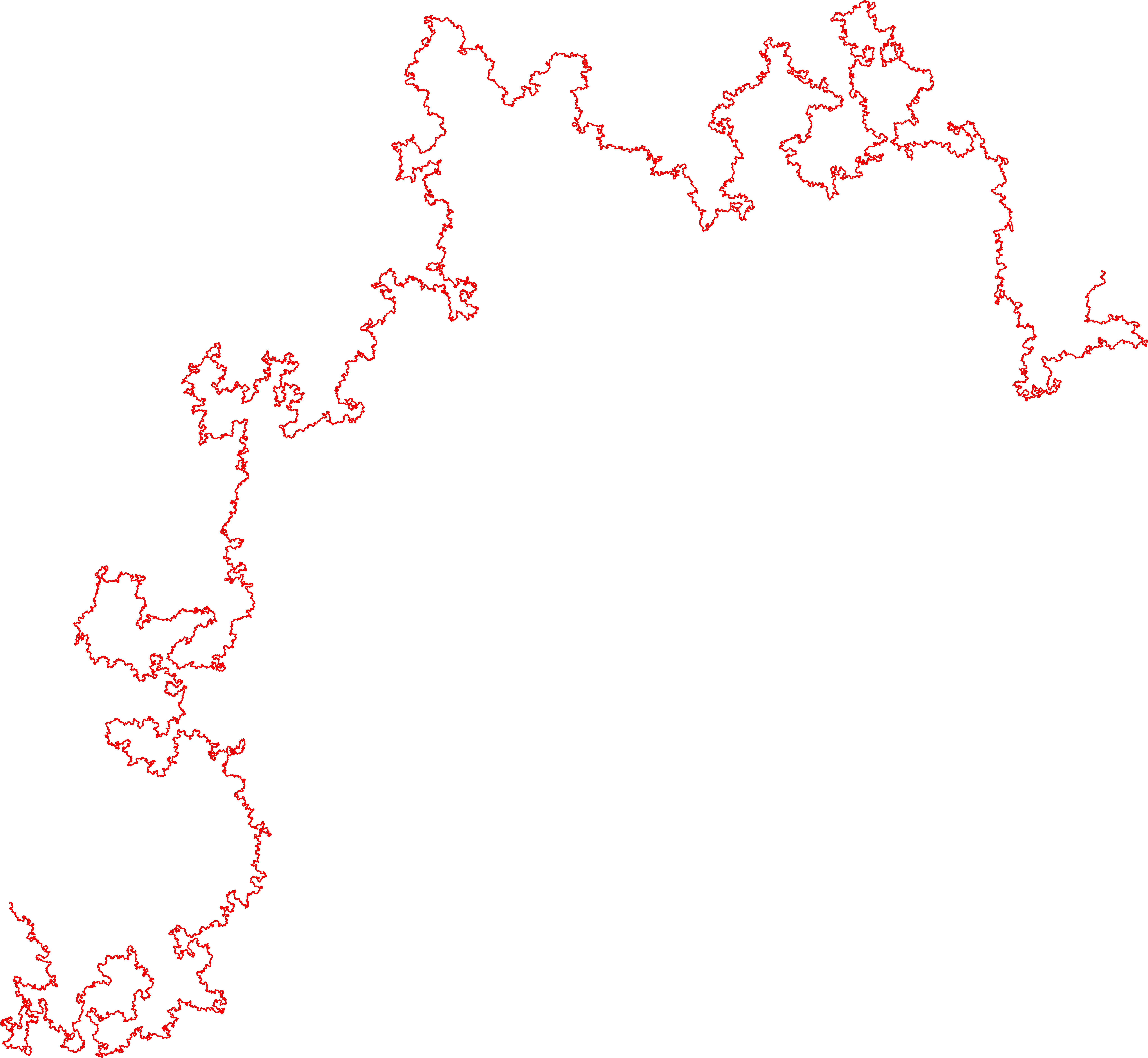}
}
\caption{Random SAWs on the square lattice
$\Z^2$ with $n=10^2$ and $n=10^8$; image  Nathan Clisby.}
\label{fig:saws}
\end{figure}

An $n$-step SAW on the integer lattice $\Z^d$ is
a map $\omega :\{0,1,\ldots,n\} \to \Z^d$, such that the Euclidean distance
between $\omega(i)$ and $\omega(i+1)$ equals $1$ (nearest-neighbour
steps), and such that $\omega(i) \neq \omega(j)$ for all $i \neq j$
(self-avoidance).
Let $\Scal_n$ denote the finite set of $n$-step SAWs
with $\omega(0)=0$ (walk starts at origin of $\Z^d$), and let $c_n$ be its
cardinality.  We declare each element of $\Scal_n$ to have equal probability,
which must therefore be $c_n^{-1}$.
Random $n$-step SAWs on the square lattice
$\Z^2$, with $n=10^2$ and $10^8$, are depicted in Figure~\ref{fig:saws}.

The $10^8$-step SAW in Figure~\ref{fig:saws}
would not be statistically
distinguishable from a SAW instead on the hexagonal lattice,
or on the triangular lattice, or indeed on any one of a wide
variety of 2-dimensional lattices.  This feature is called \emph{universality}.
It is similar to the invariance principle for Brownian motion,
which is the generalization of the central limit theorem that asserts that
(ordinary) random walk with any finite-variance step distribution
converges to Brownian motion.  The search for a corresponding statement
for SAW, i,e., the identification of a limiting probability law for SAW---a
\emph{scaling limit}---is one of the subject's big problems.
A related and in general unproven
feature is \emph{scale invariance}: a $10^{10}$-step SAW,
rescaled to the same size as the $10^8$-step SAW in Figure~\ref{fig:saws},
would be statistically indistinguishable from the $10^8$-step SAW.
The scale invariance is quantified in terms of a
universal \emph{critical exponent} whose existence has not been proven in general.

\subsubsection{The SAW connective constant}

Since $c_nc_m$ counts the number of ways that an $n$-step and an $m$-step SAW
can be concatenated, with the two subwalks possibly intersecting each other,
we have $c_{n+m} \le c_n c_m$.  From this, it readily follows
that there exists $\mu=\mu(d)$, the \emph{connective constant}, such that
$\lim_{n\to\infty}c_n^{1/n} = \mu$ and $c_n \ge \mu^n$ (see, e.g., \cite{MS93}).
Good numerical estimates and rigorous
bounds on the connective constant are known, but the exact value for $\Z^d$ is not
known for any $d \ge 2$.  For SAWs defined instead
on the hexagonal lattice, it has been proved that
$\mu=\sqrt{2+\sqrt{2}}$ \cite{D-CS12h}.
As the dimension $d$ goes to infinity, there is an asymptotic expansion
$\mu \sim 2d-1 + \sum_{n=1}^\infty a_n(2d)^{-n}$ with integer coefficients
$a_n$ whose values are known up to and including $a_{11}$ \cite{CLS07}.
The connective constant for SAWs in more general settings than $\Z^d$
is a topic of current research \cite{GL18}.

Our focus here is on
the asymptotic behaviour of the ratio $c_n/\mu^n$, which,
unlike the connective constant, is predicted to have a \emph{universal}
asymptotic behaviour.

\subsubsection{SAW critical exponents}
\label{sec:SAWexponents}

There is considerable evidence from numerical studies and from arguments
from theoretical physics that there exists $\gamma$ (depending on the
dimension $d$) such that
\begin{equation}
\label{e:cnasy}
    c_n \sim A \mu^n n^{\gamma-1} \qquad (n \to \infty).
\end{equation}
(The symbol $\sim$ denotes that the ratio of left-hand and right-hand sides has
limit $1$.)
Since $c_n \ge \mu^n$, necessarily $\gamma \ge 1$.
The \emph{susceptibility} $\chi$ is the generating function of $c_n$:
\begin{equation}
\label{e:SAWsusceptibility}
    \chi(z) = \sum_{n=0}^\infty c_n z^n
    \qquad (z \in \mathbb{C}).
\end{equation}
It has radius of convergence $z_c = \mu^{-1}$ since $c_n^{1/n}\to \mu$.
In view of \eqref{e:cnasy}, $\chi$ can be expected to obey
\begin{equation}
    \chi(z)  \sim \frac{A'}{(z_c - z)^{\gamma}}
    \qquad
    (z \uparrow z_c).
\end{equation}
Let $R_n^2$ be the average over $\Scal_n$ of $\|\omega(n)\|_2^2$.
Then $R_n$ is the root-mean-square displacement of $n$-step SAWs, which
is a measure of the average end-to-end distance of an $n$-step SAW.
There is again considerable evidence that there exists $\nu$
(depending on  $d$) such that
\begin{equation}
    R_n \sim D n^{\nu} \qquad (n \to \infty).
\end{equation}
The exponent $\nu$ quantifies scale invariance.

This article is about \emph{critical exponents} such as $\gamma,\nu$
for certain SAW and lattice spin models, with
the emphasis on $\gamma$.  There are other critical exponents that we
do not discuss.
The exponents are predicted to be universal, depending
essentially only on the dimension of the lattice.  For example, $\gamma$ and $\nu$
should have the same values on the square lattice as on the hexagonal
or triangular lattices,
unlike the connective constant.
A central problem in the subject is to prove the existence of
the critical exponents and to show that they have the values listed in
Table~\ref{table:sawexponents}.

\begin{table}[!h]
\caption{SAW critical exponents.}
\label{table:sawexponents}
\begin{tabular}{@{}rll}
\hline
$d$ & $\gamma$ & $\nu$ \\
\hline
$2$ & $\frac{43}{32}$ & $\frac 34$\\
$3$ & $1.15695300(95)$ & $0.58759700(40)$\\
$4$ & $1$ \, \text{with $\log^{1/4}$} & $\frac 12$ \, \text{with $\log^{1/8}$}\\
$\ge 5$ &  $1$ & $\frac 12$ \\
\hline
\end{tabular}
\end{table}

The rational exponents for $d=2$ in Table~\ref{table:sawexponents}
were computed by Nienhuis \cite{Nien82} using nonrigorous arguments
based on spin systems like the ones we discuss later.
An important breakthrough came with the identification of the stochastic
process
${\rm SLE}_{8/3}$ (Schramm--Loewner Evolution with parameter $\frac 83$)
as the only plausible candidate for the scaling
limit \cite{LSW04}, which additionally provided an alternate explanation for the
exponents $\frac{43}{32}$ and $\frac 34$.  However it remains an open problem to prove
the existence of the critical exponents for $d=2$,
to prove that they have the rational values
in Table~\ref{table:sawexponents},
and to prove that ${\rm SLE}_{8/3}$ truly is the scaling limit.

For $d=3$, there is no currently known stochastic process to serve as a scaling
limit for SAWs, and the best estimates for critical exponents come from
numerical work.
To compute the exponents, it is natural to attempt to enumerate SAWs
for small $n$ and then extrapolate.  Fisher and Gaunt \cite{FG64}
found $c_n$ by hand for $n \le 11$, in all dimensions.  More than half a century
later, for $d=3$ the enumeration has reached only $n=36$,
which is insufficient
for reliable high-precision estimation of the exponents.  See
Table~\ref{table:sawcount}.
It is a challenging problem in enumerative combinatorics
to produce a good algorithm to extend Table~\ref{table:sawcount} significantly.
The Monte Carlo method known as the
\emph{pivot algorithm} gives more accurate estimates for
critical exponents, and those appearing for $d=3$ in Table~\ref{table:sawexponents}
are estimates using this method \cite{CD16,Clis17}.

\begin{table}[!h]
\caption{$c_n$ for $d=3$, $n \le 36$.  The most recent values
are for $31\le n\le 36$ \cite{SBB11}.}
\label{table:sawcount}
\begin{tabular}{@{}rlrlrl}
\hline
$n$ & $c_n$ & $n$ & $c_n$ & $n$ & $c_n$\\
\hline
1 & 6
&13 & 943\,974\,510
& 25 & 116\,618\,841\,700\,433\,358\\
2 & 30
&14 & 4\,468\,911\,678
& 26 & 549\,493\,796\,867\,100\,942\\
3 & 150
&15 & 21\,175\,146\,054
& 27 & 2\,589\,874\,864\,863\,200\,574\\
4 &  726
&16 & 100\,121\,875\,974
 & 28 & 12\,198\,184\,788\,179\,866\,902\\
5 & 3\,534
&17 & 473\,730\,252\,102
 & 29 & 57\,466\,913\,094\,951\,837\,030\\
6 & 16\,926
&18 & 2\,237\,723\,684\,094
& 30 & 270\,569\,905\,525\,454\,674\,614\\
7 & 81\,390
& 19 & 10\,576\,033\,219\,614
& 31 & 1\,274\,191\,064\,726\,416\,905\,966\\
8 & 387\,966
& 20 & 49\,917\,327\,838\,734
& 32 & 5\,997\,359\,460\,809\,616\,886\,494\\
9 & 1\,853\,886
& 21 & 235\,710\,090\,502\,158
& 33 & 28\,233\,744\,272\,563\,685\,150\,118\\
10 & 8\,809\,878
& 22 &  1\,111\,781\,983\,442\,406
& 34 & 132\,853\,629\,626\,823\,234\,210\,582\\
11 & 41\,934\,150
& 23 & 5\,245\,988\,215\,191\,414
& 35 & 625\,248\,129\,452\,557\,974\,777\,990\\
12 & 198\,842\,742
& 24 & 24\,730\,180\,885\,580\,790
& 36 & 2\,941\,370\,856\,334\,701\,726\,560\,670\\
\hline
\end{tabular}
\end{table}

The exponents $\gamma=1$ and $\nu=\frac 12$
for $d\ge 5$ in Table~\ref{table:sawexponents} are the same as
those of simple random walk.
This is summarized by the statement that the \emph{upper critical dimension} is 4.
Here is an argument to guess this: Brownian paths
are 2-dimensional, and since two 2-dimensional objects generically do not intersect
in dimensions $d>4$, SAW should behave like simple random walk when $d>4$.
There is a full rigorous understanding of dimensions $d \ge 5$.  The
following theorem from \cite{HS92a} is an example of this.

\begin{theorem}
\label{thm:HS92a}
For $d \ge 5$, the scaling limit of SAW is Brownian motion, and $\gamma=1$ and $\nu= \frac 12$ in the sense that as $n \to \infty$,
\[
    c_n \sim A \mu^n ,
    \quad\quad
    R_n \sim Dn^{{1/2}} .
\]
\end{theorem}

Theorem~\ref{thm:HS92a} is proved using the lace expansion,
which was originally introduced in \cite{BS85}
and has subsequently been extended to many other high-dimensional models
including percolation \cite{Slad06,HH17book}.

For $d=4$, the logarithms in Table~\ref{table:sawexponents}
reflect the prediction that the two asymptotic formulas in
Theorem~\ref{thm:HS92a} must be modified by an additional factor $(\log n)^{1/4}$
for $c_n$ and $(\log n)^{1/8}$ for $R_n$.
We will return to such logarithmic factors
later.

For $d=2,3,4$, none of the entries in Table~\ref{table:sawexponents}
have been proved.
In 1962, Hammersley and Welsh \cite{HW62}
proved the following upper bound on $c_n$.

\begin{theorem}
\label{thm:HW}
For any $B>\pi(2/3)^{1/2}$ there
exists an $N$ such that for all $d \ge 2$,
\[
    \mu^n \le c_n \le \mu^{n} e^{B\sqrt{n}} \quad (n \ge N).
\]
\end{theorem}

Shortly thereafter, for $d \ge 3$, Kesten improved the $\sqrt{n}$ in the exponent
to $n^{2/(d+2)}\log n$ \cite{Kest64,MS93}.
For $d=2$ the best improvements since 1962 are the replacement of $B$ by $o(1)$
\cite{Hutc18}, and a proof that the upper bound holds for infinitely many $n$ when
$B\sqrt{n}$ is replaced by $n^{0.4979}$ \cite{DGHM18}.
This is slow progress in over half a century.

For $R_n$, the best results are in the following theorem.  The lower bound
was proved in \cite{Madr14} and the upper bound in \cite{D-CH13}.

\begin{theorem}
\label{thm:Rn}
For $d \ge 2$,
\[
    \tfrac{1}{6}n^{\frac{2}{3d}}  \le R_n \le o(n).
\]
\end{theorem}

The lower bound fails to prove that on average the endpoint of a SAW
is at least as far away as it is for simple random walk, namely $n^{1/2}$,
even though it
appears obvious that the self-avoidance constraint must push the SAW
farther than a walk without the constraint.  The upper bound states
that $R_n/n \to 0$ but there is no bound on the rate.  In particular,
it is not proved that there is a constant $C$ such that
$R_n  \le C n^{0.99999}$.
The large gap for $d=2,3,4$ between the predicted results
in Table~\ref{table:sawexponents} and those proven  in
Theorems~\ref{thm:HW}--\ref{thm:Rn} is
an invitation to look for more tractable models that ought to be in
the same universality class as SAW.  The weakly
self-avoiding walk is such a model.

\subsection{Weakly self-avoiding walk (WSAW)}
\label{sec:WSAW}

There are two versions of the WSAW: one based on discrete time
(also known as the \emph{Domb--Joyce model}) and one based on
continuous time (also known as the \emph{lattice Edwards model}).
Our focus is on the latter.  It differs from the SAW in two respects:
(i) the underlying simple random walk model takes its steps at random
times rather than after a fixed unit of time, and (ii) walks are allowed
to have self-intersections but are weighted as less likely according
to how much self-intersection occurs.

More precisely, let $\sigma_i$ be a sequence of independent exponential
random variables with mean $\frac{1}{2d}$.  Let $(X(t))_{t \ge 0}$ denote the
random walk on $\Z^d$ which starts at the origin at time $t=0$, waits until time
$\sigma_1$ and then steps immediately to a randomly chosen one of the $2d$ neighbours of
the origin, then waits an amount of time $\sigma_2$ until stepping to
an independently randomly chosen neighbour of its current position,
and so on.
The \emph{self-intersection local time} up to time $T$ is the random variable
\begin{equation}
    I(T)
    = \int_0^T \int_0^T \1_{X(s)=X(t)}ds \, dt
    ,
\end{equation}
which provides a measure of how much time the random walk has spent intersecting itself
by time $T$.
For fixed $g > 0$, let $c_{T,g}= E(e^{-gI(T)})$, where
$E$ denotes expectation for the random walk.  As a function of $T$,
$c_{T,g}$ is analogous to the sequence $c_n$ for SAW.
Every walk contributes to $c_{T,g}$, but an exponential weight diminishes the
role of walks with large self-intersection local time.
The elementary argument which led to the existence of the connective constant
generalizes to $c_{T,g}$, and yields the conclusion that there exists
$\nu_c(g) \le 0$ such that $\lim_{T\to\infty}c_{T,g}^{1/T} = e^{\nu_c}$
and $c_{T,g} \ge e^{\nu_c T}$.  Thus the \emph{susceptibility}
\begin{equation}
    \chi(g,\nu)  = \int_0^\infty
    c_{T,g} e^{-\nu T}
    dT \qquad (\nu \in \mathbb{R})
\end{equation}
is finite if and only if $\nu > \nu_c$.

WSAW is predicted to be in the same universality class as SAW for all $g>0$,
meaning that it has the same critical exponents and scaling limits as SAW.
The following theorem is an example of this, for the upper critical
dimension $d=4$ and for sufficiently
small $g>0$ \cite{BBS-saw4-log}.
It reveals that $\gamma=1$ with a modification by a
logarithmic correction as indicated in Table~\ref{table:sawexponents}.
In the physics literature, the computation of logarithmic corrections for $d=4$ goes
back half a century \cite{LK69,BGZ73,WR73,Dupl86}.
A number of related results have been proved for the 4-dimensional WSAW
\cite{BBS-saw4,ST-phi4,BSTW-clp},
all of which are consistent with the predictions for SAW.

\begin{theorem}
\label{thm:saw4-log}
For $d = 4$ and small $g>0$, as $t=\nu-\nu_c \downarrow 0$,
\begin{align*}
    \chi(g,\nu)
    & \sim A_g\frac {1}{t} |\log t|^{ \frac 14}
    .
\end{align*}
\end{theorem}

The proof of Theorem~\ref{thm:saw4-log} is
based on a rigorous and nonperturbative implementation of the renormalization
group approach \cite{WK74}.
The renormalization group has for decades been one of the basic tools of theoretical
physics, for which  Wilson
was awarded the Nobel Prize in Physics in 1982.
Its reach extends across critical phenomena, many-body theory,
and quantum field theory.  We make no attempt to refer to the vast
physics literature, e.g., \cite{Card96}.

In a 1972 paper with the intriguing title
``Critical exponents in $3.99$ dimensions'' \cite{WF72},
Wilson and Fisher considered the dimension $d$ as a
\emph{continuous} variable $d=4-\epsilon$,
and applied the renormalization group approach to compute critical exponents in dimension $4-\epsilon$ for small $\epsilon>0$.  This captures the idea that
the critical behaviour can be expected
to vary in a continuous manner as the dimension varies,
so dimensions below $d=4$ can be regarded as a perturbation of $d=4$.
Within physics, this has become well developed and it is found that, e.g.,
${ \gamma = 1 + \frac 18 \epsilon + \cdots + ({\rm known})\epsilon^6 + \cdots}$.
Although presumably a divergent asymptotic expansion,
such $\epsilon$-\emph{expansions} have been used to obtain numerical estimates
of critical exponents for $d=3$.
However, from the perspective of mathematics, the dimension is not a continuous
variable and this raises more questions than it answers.

\subsection{Long-range walks}

A different idea to move slightly below the upper critical dimension
was also proposed in 1972 \cite{FMN72,SYI72}.
In this framework, the upper critical dimension (formerly $d=4$) assumes
a continuous value $d_c \in (0,4)$.  In particular,
 for $d=1,2,3$ we can choose
$d_c=d+\epsilon$, and thereby study integer dimension
$d$ below $d_c$ without the need to define the WSAW in any non-integer dimension.
In our present context, this idea can be formulated in terms of walks
taking long-range steps, as follows.

\begin{figure}[!h]
\hspace{10mm}
\centering
\includegraphics[scale = 0.18]{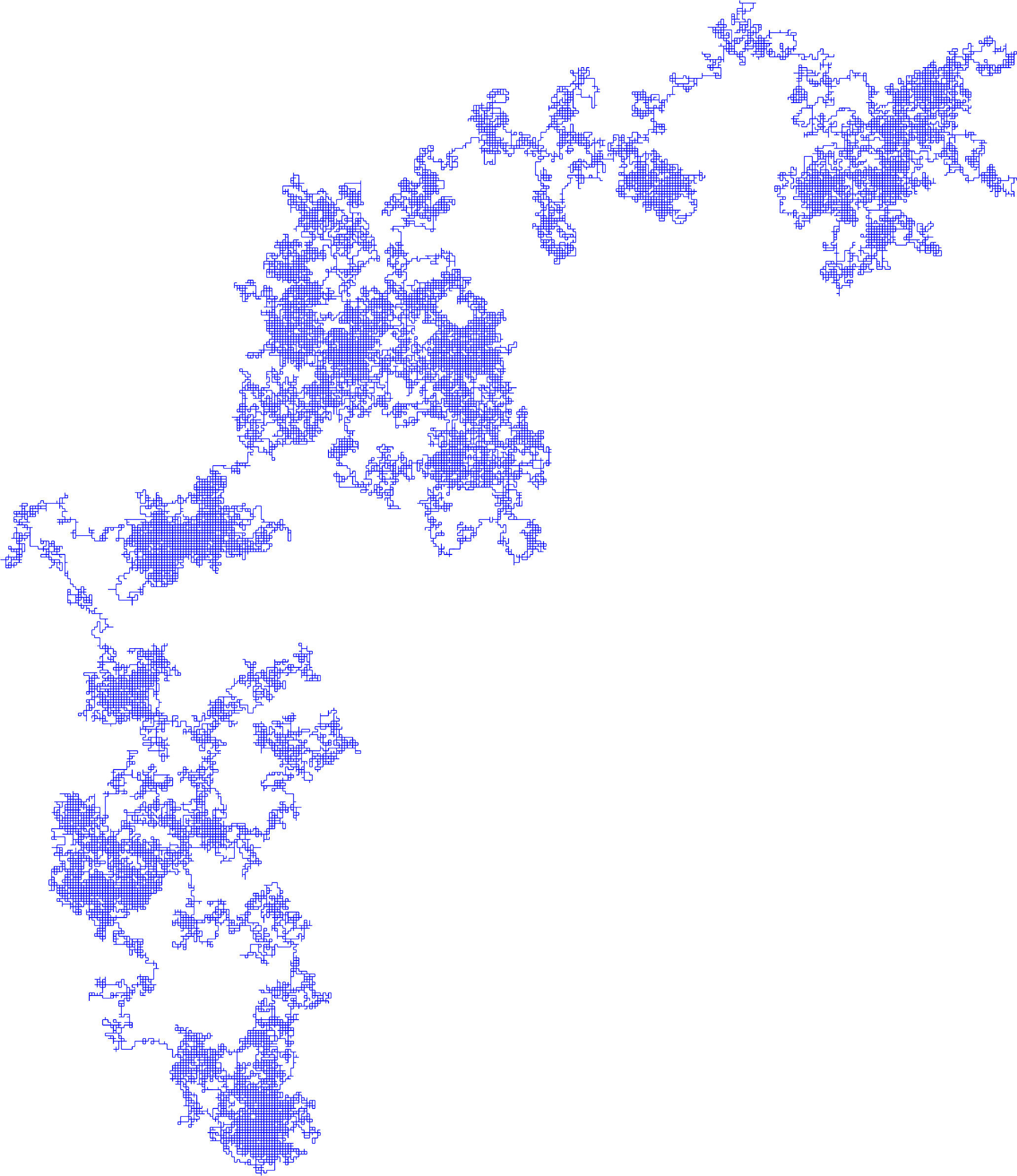}
\hspace{10mm}
\includegraphics[width=50mm]{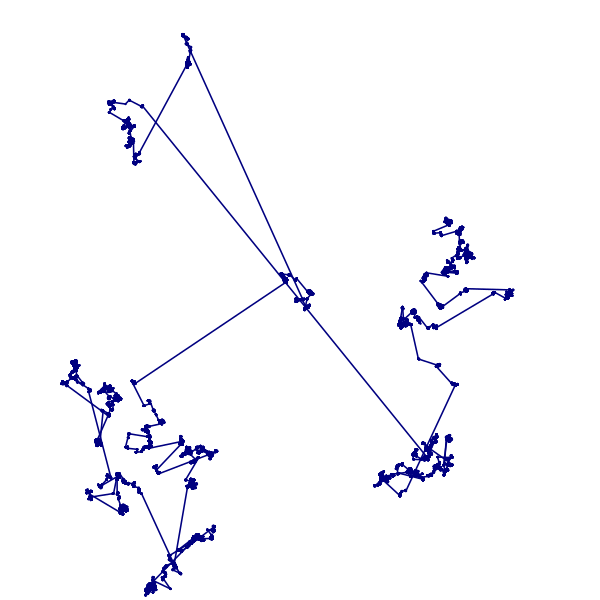}
\caption{$2$-dimensional $10^5$-step nearest-neighbour (left) and long-range
(right, $\alpha = 1.1$) walks (not to scale: the diameter of the left walk is
about 100 times smaller than that of the right walk); image Nathan Clisby. }
\label{fig:long-range-srw}
\end{figure}

The long-range steps are defined in terms of a parameter
$\alpha \in (0,2)$.
Let $d=1,2,3$, and
consider the random walk on ${\mathbb Z}^d$ that takes independent steps of length $r$ (in any direction) with probability proportional to
$r^{-(d+\alpha)}$.
This step distribution has infinite variance, a \emph{heavy tail}.
A convenient choice of such a step distribution is the
fractional power $-(-\Delta)^{\alpha/2}$ of the discrete Laplace operator
\begin{equation}
\label{e:Deltadef}
    (\Delta f)_x = \sum_{e\in \Z^d:\|e\|_2=1} (f_{x+e}-f_x).
\end{equation}
Thus we consider the random walk on $\Z^d$ with transition probabilities
\begin{equation}
\label{e:pxy}
    p_{x,y}= {\mathbb P}(\text{{ next step to $y$}}|\text{{ now at $x$}})
    \propto   -((-\Delta)^{\alpha/2})_{x,y} \asymp \frac{1}{\|x-y\|_2^{d+\alpha}}
\end{equation}
(the notation $f\asymp g$ means $cg \le f \le Cg$ for some constants $c,C$).
This heavy-tailed random walk converges to an $\alpha$-stable
process.  The paths of an $\alpha$-stable process have dimension $\alpha$
\cite{BG60}, so two such paths
generically do not intersect in dimensions $d>2\alpha$.  This
suggests that in dimensions $d>d_c=2\alpha$,  long-range SAW or WSAW
should behave like the $\alpha$-stable process.  Figure~\ref{fig:long-range-srw}
shows a 2-dimensional long-range simple random walk with $\alpha = 1.1$,
next to a nearest-neighbour walk for comparison.
The heavy tail of the long-range
walk produces big jumps, which in turn create fewer self intersections,
thereby making it easier for a walk to be self-avoiding and
lowering the upper critical dimension.

\vspace*{-15pt}

We can define a long-range model of SAW as follows.  A
long-range $n$-step SAW is any sequence
$\omega=(\omega(0),\ldots, \omega(n))$ with $\omega(i) \in \Z^d$
and $\omega(i) \neq \omega(j)$ for $i \neq j$.  Let $\omega(0)=0$.
The probability of $\omega$ is the product
$\prod_{i=1}^n p_{\omega(i-1),\omega(i)}$, with $p_{x,y}$ given
by \eqref{e:pxy}.
The following theorem \cite{Heyd11} proves that
this SAW does behave like the unconstrained long-range random walk in dimensions $d\ge 1$
as long as $\alpha < \frac d2$.  This is a long-range version
of Theorem~\ref{thm:HS92a}; its proof is also based on the lace expansion.
A technical point is that the theorem actually applies to a so-called
\emph{spread-out} version of the long-range SAW, a small modification.

\begin{theorem}
\label{thm:SAWlr}
For $\alpha \in (0,2)$ and $d>2\alpha$, the scaling limit of
spread-out long-range SAW is an $\alpha$-stable process, and the
critical exponents are $\gamma=1$, $\nu = \frac 1\alpha$.
\end{theorem}

However, our primary interest here is to go below $d_c$ to observe scaling
behaviour that is different from that of the $\alpha$-stable process.
For this, we consider WSAW and its susceptibility $\chi$ defined as in
Section~\ref{sec:SAW}\ref{sec:WSAW}
but with the expectation $E$ now with respect to the
continuous-time long-range random walk.
We choose $\alpha = \frac 12 (d+\epsilon)$, so that $d=d_c-\epsilon$ is
below the critical dimension $d_c=2\alpha$.
The following theorem \cite{Slad18} gives an example of an $\epsilon$-expansion.  It is proved using a rigorous renormalization group method.
The restriction on $g$ in the hypothesis of the theorem is
used in the proof,
but the statement is expected to be true for all $g>0$.
Further results are obtained in \cite{LSW17}.
A related paper which is focused on renormalization rather than critical
exponents is \cite{MS08}.

\begin{theorem}
\label{thm:WSAWlr}
Let $d = 1,2,3$.
For small $\epsilon>0$, for $\alpha = \frac 12 (d+\epsilon)$,
and
for $g\in [c\epsilon,c'\epsilon]$ for some $c<c'$,
there is a constant $C$ such that
as $t=\nu-\nu_c \downarrow 0$,
\begin{align*}
    &
    C^{-1}\frac{1}{t^{ 1+ \frac{1}{4} \frac\epsilon\alpha -C\epsilon^2}}
    \le
    \chi(g,\nu) \le C\frac{1}{t^{ 1+ \frac{1}{4}\frac\epsilon\alpha +C\epsilon^2}}
    ,
    \quad \text{i.e.,} \;\;
    { \gamma  = 1 + \frac{1}{4} \frac\epsilon\alpha + O(\epsilon^2)}.
\end{align*}
\end{theorem}

\section{Spin systems}

Spin systems are basic models in statistical mechanics.
We discuss two examples here: the Ising and $|\varphi|^4$ models.
At first sight, spin
systems appear to be unrelated to SAW, but a connection will
be made in Section~\ref{sec:n0}.

\subsection{Ising model}

\begin{figure}[h]
\begin{center}
\includegraphics[scale = 0.35]{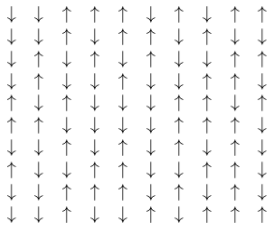}
\end{center}
\caption{
An Ising spin configuration.}
\label{fig:ising-arrows}
\end{figure}

The most fundamental spin system is the Ising model
of ferromagnetism, which is defined
as follows.  Let $\Lambda \subset \Z^d$ be a finite box.
An \emph{Ising spin configuration} on $\Lambda$ is an assignment of $+1$ or
$-1$ to each site in $\Lambda$, i.e., $\sigma = (\sigma_x)_{x\in\Lambda}$
with $\sigma_x \in \{-1,1\}$.  An example for $d=2$ is depicted in
Figure~\ref{fig:ising-arrows}.

Spin configurations are random, with a probability distribution parametrized
by \emph{temperature} $T$ and determined by the
energy of $\sigma$ which is defined to be
\begin{equation}
    H_{\Lambda}(\sigma)
    =
    \frac 12 \sum_{x\in\Lambda}  \sigma_x (-\Delta \sigma)_x .
\end{equation}
Here $\Delta$ is the discrete Laplace operator \eqref{e:Deltadef}, restricted to $\Lambda$.
Apart from an unimportant constant, $H_{\Lambda}(\sigma)$ is equal to
$-  \sum_{x\sim y} \sigma_x\sigma_y$ where the sum is over all pairs of neighbouring
sites in $\Lambda$.
At temperature $T$, the
probability of $\sigma$ is given by the \emph{Boltzmann weight}
\begin{equation}
    P_{T,\Lambda}(\sigma)
    =
    \frac{e^{-\frac 1T H_{\Lambda}(\sigma) }}
    {\sum_{\sigma}e^{-\frac 1T H_{\Lambda}(\sigma) }}
    \propto
    e^{\frac{1}{T} \sum_{x\sim y} \sigma_x\sigma_y}
    .
\end{equation}
Thus spin configurations with more alignment between neighbouring spins
are more likely than those with less alignment, and this effect is magnified
for small $T$ compared to large $T$.

\begin{figure}[h]
\begin{center}
\includegraphics[width = .20 \textwidth]{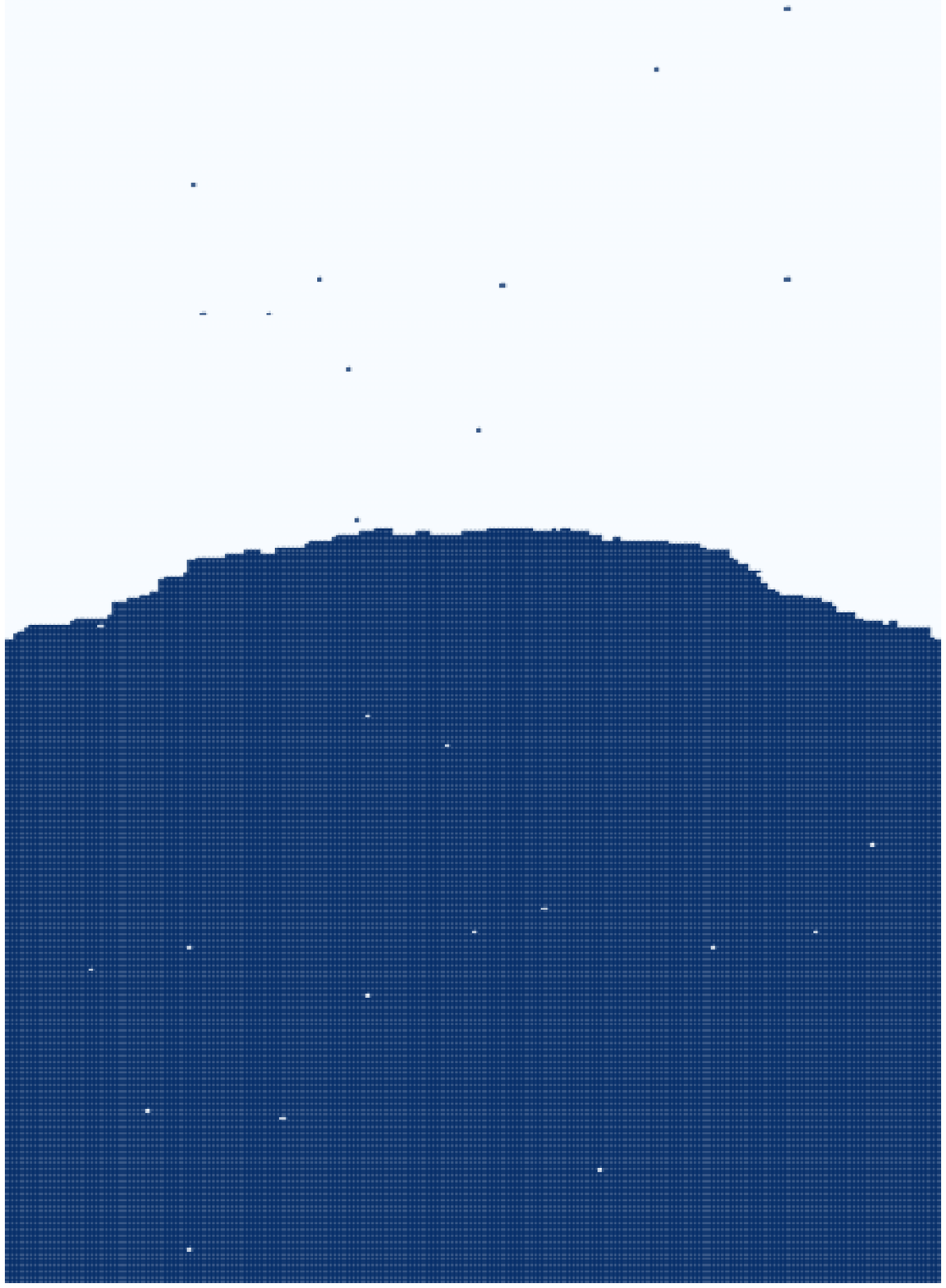}%
\includegraphics[width = .20 \textwidth]{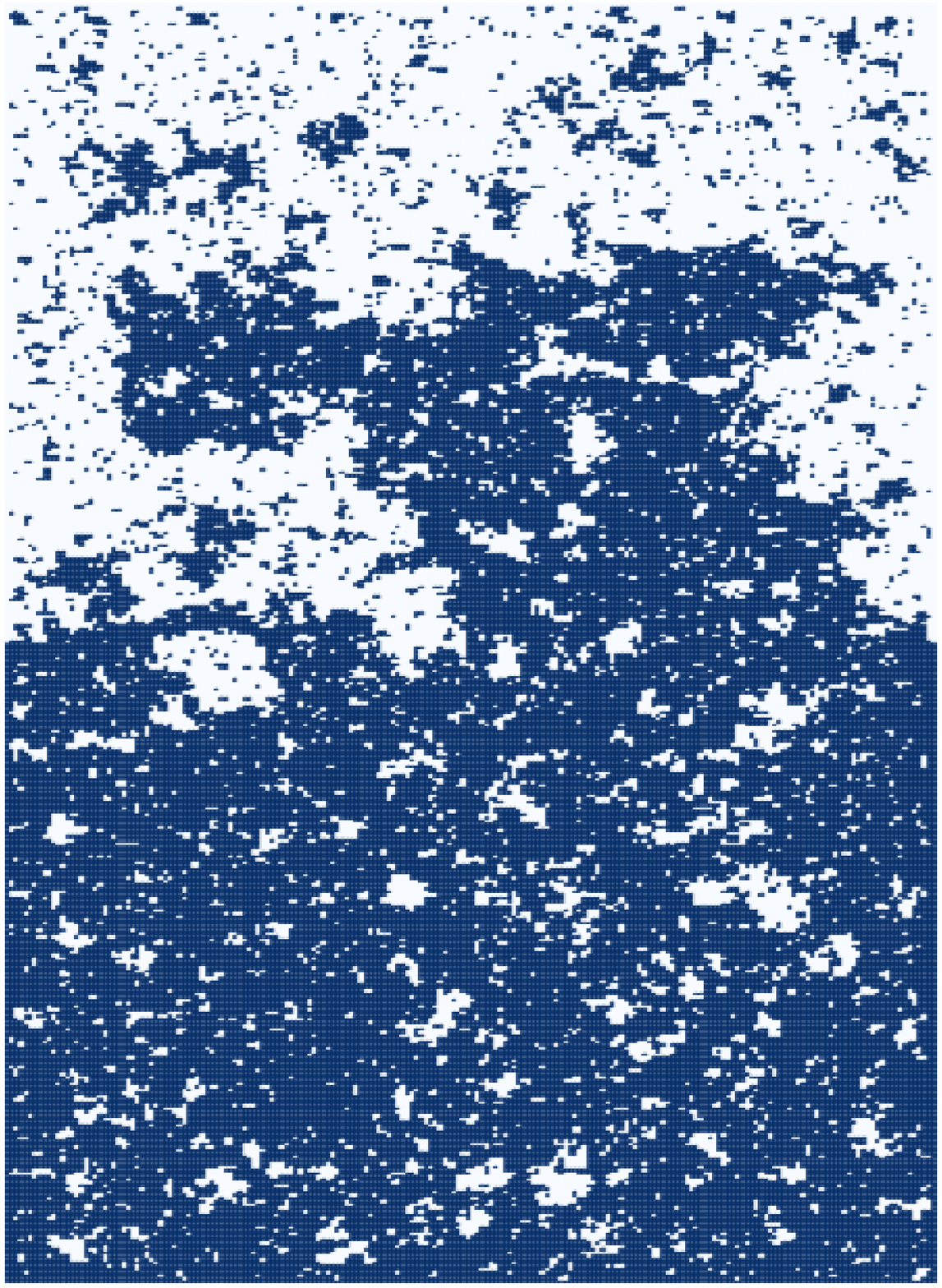}%
\includegraphics[width = .20 \textwidth]{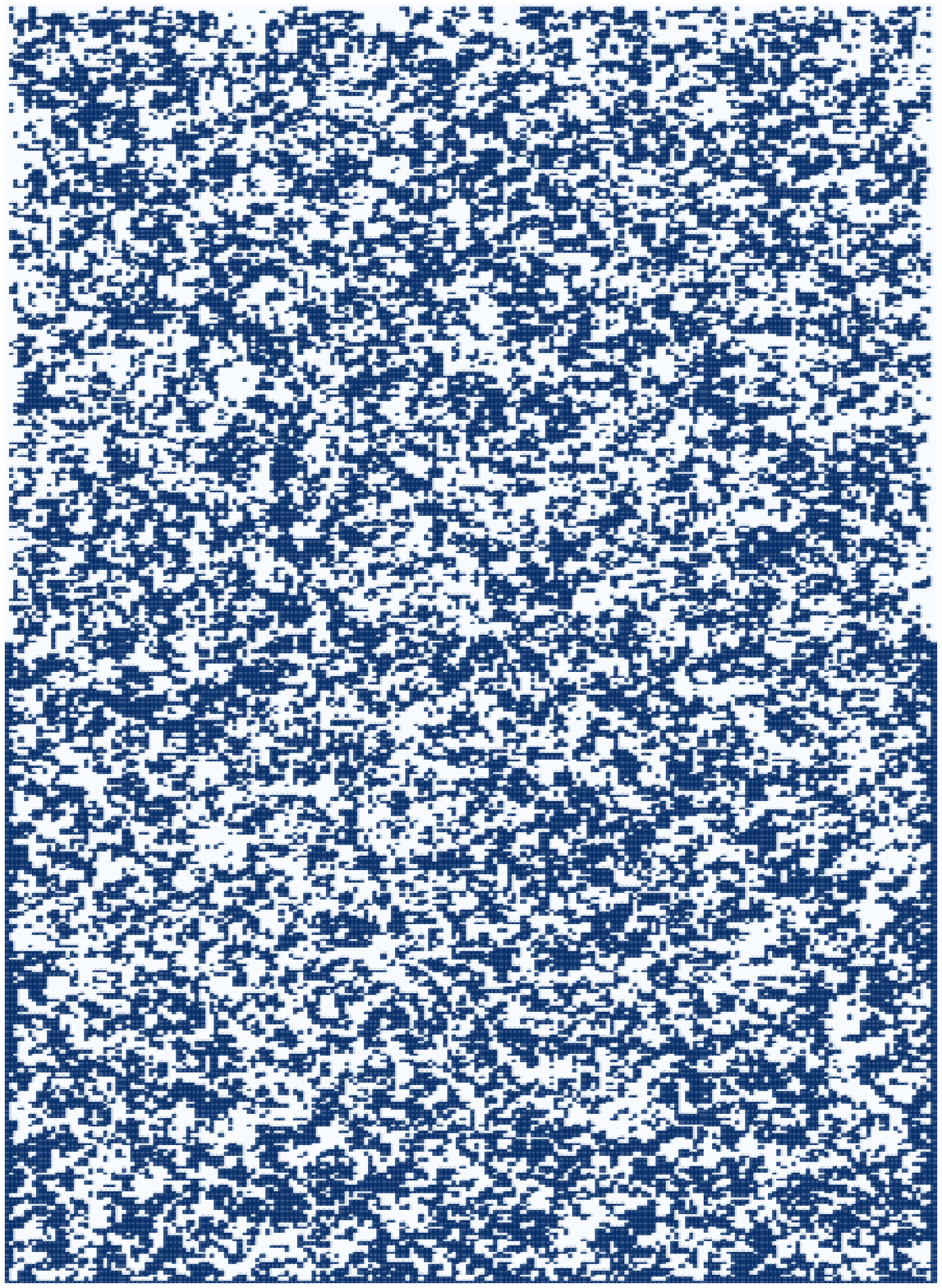}%
\end{center}
\begin{center}
{\scriptsize Low temperature $T<T_c$ \hspace{2mm} Critical temperature $T=T_c$ \hspace{3mm}
High temperature $T>T_c$}
\end{center}
\caption{Ising configurations on a $200 \times 400$ box,
with boundary spins fixed white on top half and dark on bottom half.
}
\label{fig:ising-2D}%
\end{figure}

For dimensions $d \ge 2$, there is a
\emph{critical} temperature $T_c$ such that when $T>T_c$ typical spin
configurations are disordered, whereas for $T<T_c$ there is long-range order.
This is depicted in Figure~\ref{fig:ising-2D} where the $+/-$ symmetry
is broken by a boundary condition.  The behaviour at $T_c$, and
as $T$ approaches $T_c$, is of great current interest and there
is a vast literature, particularly for $d=2$ where the model is exactly solvable
and exciting connections with ${\rm SLE}$ have been discovered, e.g.,
\cite{CDHKS14}.  At the critical temperature, the rich geometric structure
apparent in Figure~\ref{fig:ising-2D} is scale invariant.
Critical exponents are rigorously known for $d=2$ and for $d>4$ but not for
$d=3$, although in the physics literature the conformal
bootstrap has been used to compute exponents to high accuracy for $d=3$
\cite{EPPRSV14}.
A recent survey of mathematical work on the Ising and related models can be
found in \cite{Dumi19}.

\subsection{$|\varphi|^4$ model}

\begin{figure}[h]
\begin{center}
\scalebox{0.75}
    {\input{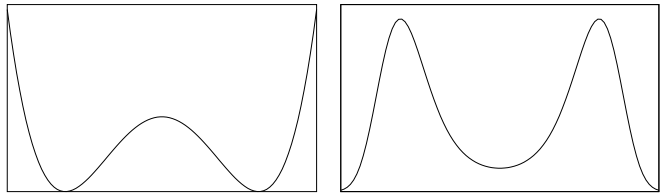}}
\end{center}
\caption{\label{fig:V}
For $n=1$, the double-well potential $V$ (left) and single-spin density
$e^{-V}$ (right).}
\end{figure}

The $|\varphi|^4$ model is an extension of the Ising model in which the
Ising spin $\sigma_x \in \{-1,+1\}$ is replaced by an $n$-component vector
spin $\varphi_x \in \R^n$ ($n\ge 1$).
To preserve translation invariance we replace the box used for the Ising
model by one with periodic boundary conditions, i.e., $\Lambda$ is a discrete
$d$-dimensional torus.
The {\it a priori} or \emph{single-spin} distribution
of $\varphi_x$ is set to be proportional to $e^{-V(\varphi_x)} d\varphi_x$, where
$d\varphi_x$ is Lebesgue measure on $\R^n$ and
\begin{equation}
     V(\varphi_x)
     = \tfrac{1}{4}g |\varphi_x|^4 + \tfrac{1}{2} \nu |\varphi_x|^2
\end{equation}
with $g>0$, $\nu\in \R$, and with $|\varphi_x|$ the Euclidean norm of $\varphi_x\in
\R^n$.  We are primarily interested in
$\nu<0$, in which case for $n=1$ the potential $V$ has the double-well shape
of Figure~\ref{fig:V}.
The probability density of a spin configuration
$(\varphi_x)_{x\in \Lambda}\in (\R^n)^{|\Lambda|}$ is then proportional to
the Boltzmann weight
\begin{align}
\label{e:phi4}
     dP_{g,\nu,\Lambda}(\varphi)
     & \propto
     e^{
      -
     \sum_{x \in \Lambda} (
      V(\varphi_x)
      +
      \tfrac{1}{2} \varphi_x\cdot (-\Delta \varphi)_x
      )
     }
     \prod_{x\in \Lambda}
     d\varphi_x
     .
\end{align}

For $n=1$, spins are more likely to assume values near one of the two minima
of the double well.  For $n >1$, there is a continuous set of minima.
The Laplacian term in \eqref{e:phi4} discourages large differences between
neighbouring spins and is thus a ferromagnetic interaction.
For example, for $n=1$ it encourages
the spins to break the symmetry and primarily prefer one minimum over the other.
Now $\nu$ plays the role played by the temperature $T$ in the Ising model, and there
is a phase transition and corresponding critical exponents associated with
a critical value $\nu_c(g)<0$.
For $\nu<\nu_c$, spins are typically aligned, whereas they are disordered
for $\nu>\nu_c$.
The existence of a phase transition is proved for $d \ge 3$ for
general $n \ge 1$ in \cite{FSS76}, and for $d=2$ and $n=1$ in \cite{GJS75};
the Mermin--Wagner theorem states that there is no phase transition for $n >1$
when $d=2$.

The \emph{susceptibility} is defined by
\begin{equation}
\label{e:phi4susceptibility}
    \chi(g,\nu) = \lim_{\Lambda \uparrow \Z^d} \frac 1n \sum_{x \in \Lambda}
    \int_{(\R^n)^{|\Lambda|}} \varphi_0 \cdot \varphi_x
    \;
    dP_{g,\nu,\Lambda}(\varphi),
\end{equation}
assuming the limit exists.
It represents the sum over all $x$ of the correlation of the spin at $0$
with the spin at $x$.  If $\nu$ is above the critical point $\nu_c$ then
correlations remain summable, but there is divergence at $\nu=\nu_c$.
The predicted behaviour of the susceptibility,
as $t =\nu - \nu_c \downarrow 0$, is
\begin{equation}
    \chi(g,\nu) \sim A_{g,n}  \frac{1}{t^{ \gamma}}
    ,
\end{equation}
with a universal critical exponent $\gamma$ (depending on $d,n$, but not $g$), and
with a logarithmic correction for $d=4$.  It was proven in 1982 that $\gamma=1$
for $d>4$ \cite{Aize82,Froh82}.

The concept of universality was discussed in
Section~\ref{sec:SAW}\ref{sec:SSAW}.
It is predicted that the 1-component $|\varphi|^4$ model is in the
same universality class as the Ising model, and that more generally the
$n$-component $|\varphi|^4$ model lies in the universality class of the
model in which the single spin distribution $e^{-V(\varphi_x)}d\varphi_x$
is replaced by the uniform distribution on the sphere
of radius $\sqrt{n}$ in $\R^n$.
In addition, if the nearest-neighbour interaction
given by the Laplacian is replaced by any other finite-range interaction
respecting the symmetries of $\Z^d$, then the resulting model is predicted
to be in the
same universality class as the nearest-neighbour model.

The following theorem from \cite{BBS-phi4-log} determines the
asymptotic form of the susceptibility for $d=4$.
Its proof is via a rigorous renormalization group method.
It and related work \cite{ST-phi4,BSTW-clp} give
extensions of mathematical work from the 1980s \cite{GK85,HT87,FMRS87}.
(A caveat for Theorems~\ref{thm:phi4-log}--\ref{thm:phi4lr}
is that the susceptibility is defined with the
infinite volume limit taken through a sequence of tori of period $L^N$
with fixed large $L$, as $N \to \infty$.)

\begin{theorem}
\label{thm:phi4-log}
For $d=4$, $n \ge 1$, small $g>0$, as $t=\nu-\nu_c \downarrow 0$,
\begin{align*}
    \chi(g,\nu) &\sim  A_{g,n} \frac{1}{t}|\log t|^{\frac{n+2}{n+8}}.
\end{align*}
\end{theorem}

To go below the upper critical dimension, we again consider a long-range
version of the model, by replacing  the Laplacian term
$\varphi_x\cdot (-\Delta \varphi)_x$ in \eqref{e:phi4} by a term
$\varphi_x\cdot ((-\Delta)^{\alpha/2} \varphi)_x$ with
fractional Laplacian and $\alpha \in (0,2)$.
The upper critical dimension is again $d_c=2\alpha$.
Several rigorous results use the lace expansion to
prove mean-field behaviour for various long-range models when $d>d_c$,
e.g., \cite{HHS08} for the Ising model.
The following theorem from \cite{Slad18}
concerns dimensions $d=d_c-\epsilon$ which lie slightly below $d_c=2\alpha$.
Related results are proved in \cite{LSW17}, and earlier mathematical papers for
long-range models are \cite{BMS03,Abde07,ACG13}.

\begin{theorem}
\label{thm:phi4lr}
Let $d=1,2,3$.
For $n \ge 1$, for small $\epsilon>0$, for $\alpha = \frac 12 (d+\epsilon)$,
and for $g\in [c\epsilon,c'\epsilon]$ for some $c<c'$,
there is a constant $C$ such that
as $t=\nu-\nu_c \downarrow 0$,
\begin{align*}
    &
    C^{-1}
   \frac{ 1}{t^{ 1+ \frac{n+2}{n+8} \frac\epsilon\alpha -C\epsilon^2}}
    \le
    \chi(g,\nu) \le
    C\frac{1}{t^{ 1+ \frac{n+2}{n+8} \frac\epsilon\alpha +C\epsilon^2}},
    \quad i.e.,\;
    { \gamma  = 1 + \frac{n+2}{n+8} \frac\epsilon\alpha + O(\epsilon^2)}.
\end{align*}
\end{theorem}

\section{Supersymmetry and $n=0$}
\label{sec:n0}

Comparison of Theorems~\ref{thm:phi4-log}--\ref{thm:phi4lr} with
Theorems~\ref{thm:saw4-log} and \ref{thm:WSAWlr} reveals that when
the $|\varphi|^4$ theorems have the number $n$ of components replaced
by $n=0$ then the WSAW statements result.
This apparently curious observation is not an accident.

Indeed, de Gennes argued in 1972 \cite{Genn72}
that the
self-avoiding walk model \emph{is} the $n=0$ version of the
$n$-component spin model.
Roughly speaking, his reasoning was that
the susceptibility of an $n$-component spin model
has a geometric representation involving a self-avoiding walk and loops, with
the loops weighted by $n$.  When $n$ is set equal to zero, only the SAW remains.

This has been a very productive observation in physics.  For example, once
it has been established that the $|\varphi|^4$ exponent is
$\gamma = 1 + \frac{n+2}{n+8} \frac{\epsilon}{\alpha}+\cdots$,
then the inference is made that the SAW exponent is $\gamma =
 1 + \frac{1}{4}\frac{\epsilon}{\alpha} +\cdots$.
However, from a mathematical perspective, just as it halted progress to
consider non-integer dimensions $d=4-\epsilon$, it is also
problematic to contemplate
the notion of a $0$-component spin, or
of a limit $n \downarrow 0$ when the dimension $n$ of the spin
is a natural number.

An alternate idea from physics with a similar conclusion to de Gennes's
was proposed independently in 1980 by
Parisi and Sourlas \cite{PS80} and by McKane \cite{McKa80}.
Their idea was that while
an $n${-component boson field} $\varphi$ (usual spin)
contributes a factor $n$ for
every loop in the geometric representation of the susceptibility,
an $n$-component \emph{fermion} field contributes $-n$.
When combined, all loops cancel, leaving the self-avoiding walk.
From a mathematical point of view,
this realization of zero components as $n-n$ is less problematic
than setting $n=0$ or considering the limit $n \downarrow 0$,
and it leads to a theorem.  Some history of the mathematical work
in this direction can be found in \cite{BBS-brief}.  An important
early step was \cite{BM91}, which was inspired by \cite{Lutt83}.

Fermion fields are often defined in terms of Grassmann variables, which
multiply with an
anti-commuting product. A fermion field can also be constructed
using differential forms with their anti-commuting wedge product,
and we follow this route in the following.

Given any finite set $\Lambda$ of cardinality $M=|\Lambda|$, we consider
$2M$ real coordinates and corresponding 1-forms:
\begin{equation}
    u_1,v_1, \ldots , u_{M},v_{M}
    \quad
    \text{and}
    \quad
    du_1,dv_1,\ldots,du_{M},dv_{M}.
\end{equation}
The wedge product $\wedge$ is associative and anti-commuting, e.g.,
$du_x \wedge dv_y = - dv_y \wedge du_x$.
Let $u=(u_1,\ldots,u_M)$ and similarly for $v$.
A \emph{form} is a function of $(u,v)$ times a product
of 1-forms, or a linear combination of these.
A sum of forms which each contains a product of $p$ distinct 1-forms is
called a $p$-form.  Due to the anti-commutativity, $p$-forms are zero if
$p>2M$.  Also, any $2M$-form $F$ can be written uniquely as
 $F=f(u,v) du_1 \wedge dv_1 \wedge \cdots \wedge du_{M} \wedge dv_{M}$.  Integration of a $2M$-form $F$ is defined by
 the Lebesgue integral
 \begin{equation}
\label{e:intdef}
    \int F \;\;  =
    \int_{\R^{2{M}}} f(u,v) du_1dv_1\cdots du_{M}dv_{M}
    ,
\end{equation}
and the integral of a $p$-form is defined to be zero if $p<2M$.
This definition of integration extends by linearity to arbitrary forms.

We write the $2M$ real coordinates in terms of $M$ complex coordinates:
\begin{equation}
    \phi_x = u_x+iv_x, \quad \bar\phi_x = u_x-iv_x, \quad
    d\phi_x = du_x+idv_x, \quad d\bar\phi_x = du_x-idv_x.
\end{equation}
Let $\psi_x = \frac{1}{\sqrt{2\pi i}} d\phi_x$ and
$\bar\psi_x = \frac{1}{\sqrt{2\pi i}} d\bar\phi_x$ .
The field $\phi_x$ is a 2-component \emph{boson} field on $\Lambda$, and $\psi_x$ is
a 2-component \emph{fermion} field.
We define the differential forms
\begin{equation}
\label{e:taudef}
    \tau_x = \phi_x\bar\phi_x +
    \psi_x \wedge \bar\psi_x,
    \quad
    \tau_{\Delta,x} = \phi_x(-\Delta \bar\phi)_x
    + \psi_x  \wedge (-\Delta \bar\psi)_x.
\end{equation}
Smooth functions of forms are defined by Taylor expansion in $\psi,\bar\psi$,
which terminates as a Taylor polynomial due to the anti-commutativity.
For example,
\begin{equation}
    e^{-\sum_{x\in \Lambda} \tau_x} = e^{-\sum_{x\in \Lambda} \phi_x\bar\phi_x}\sum_{m=0}^{M} \frac{(-1)^m}{m!}
    \Big(\sum_{x\in \Lambda} \psi_x \wedge \bar\psi_x\Big)^m.
\end{equation}

The susceptibility of the WSAW on a finite subset $\Lambda \subset \Z^d$
is then given by the remarkable identity
\begin{equation}
\label{e:intrep}
    \chi_\Lambda(g,\nu)  =
    \sum_{x\in \Lambda}
    \int_{\R^{2|\Lambda|}}
    e^{-\sum_{z\in\Lambda}(g\tau_z^2 + \nu \tau_z + \tau_{\Delta,z})}\bar\phi_0\phi_x
    ,
\end{equation}
with the integral on the right-hand side evaluated according to the definition
of the integral as in \eqref{e:intdef} after conversion of the complex coordinates
to real coordinates \cite{BM91}.
The identity \eqref{e:intrep} is discussed in
detail in \cite{BBS-brief}, where a proof is given based on
\emph{supersymmetry}, which is a symmetry
that relates the boson and fermion fields.  Replacement of
$-\Delta$ by $(-\Delta)^{\alpha/2}$ in the definition of $\tau_{\Delta,x}$
in \eqref{e:taudef} and \eqref{e:intrep}
gives a corresponding identity for the long-range model.

The right-hand side of \eqref{e:intrep} is reminiscent of the
right-hand side of the definition of the susceptibility in
\eqref{e:phi4susceptibility}.  For example, the bosonic part of the exponent
on the right-hand side of \eqref{e:intrep}
matches the exponent on the
right-hand side of the Boltzmann weight \eqref{e:phi4} for the $|\varphi|^4$ model.
The renormalization group method discussed in Section~\ref{sec:RG}
applies equally well
with or without the presence of the fermion field.
This allows a treatment of
WSAW simultaneously with the $n$-component $|\varphi|^4$ model,
as the $n=2-2 = 0$ case, and provides a mathematically
rigorous implementation of de Gennes's idea,
via the supersymmetric formulation introduced by Parisi and Sourlas and by McKane.

\section{Renormalization group (RG) method}
\label{sec:RG}

Theorems~\ref{thm:saw4-log}, \ref{thm:WSAWlr}
and \ref{thm:phi4-log}--\ref{thm:phi4lr} are proved via a rigorous
RG method.  Aspects of the RG method are described in this section.

\subsection{RG Strategy}

Scaling limits in critical phenomena have the feature of scale invariance
visible in the long SAW in Figure~\ref{fig:saws}
and in the simulation of the critical Ising model
in Figure~\ref{fig:ising-2D}.
Wilson's brilliant strategy to exploit the scale invariance to simultaneously
explain universality and provide a practical tool for the computation of
universal quantities such as critical exponents
can be outlined schematically as follows:
\begin{enumerate}
\item
Introduce a mapping, the RG map, that maps a model at one scale to a model
at a larger scale.
Scale invariance corresponds to a fixed point of the RG map.
\item
A stable fixed point has a domain of attraction under iteration of the RG map.
The domain of attraction is a universality class of models.
\item
The universal
properties of a scale invariant model can be calculated from
the behaviour of
the RG map in the vicinity of the fixed point.
\end{enumerate}

The ``group'' operation in the term ``renormalization group'' is the
operation of composition of maps.  The maps are generally not invertible,
so this is a semigroup with identity rather than a group.  The terminology
renormalization ``group'' has nevertheless become commonplace.

The proofs of Theorems~\ref{thm:phi4-log}--\ref{thm:phi4lr}
for the $|\varphi|^4$ model are based on
the above strategy, with some adaptation due to lattice effects.
As discussed in Section~\ref{sec:n0}, the proofs of
Theorems~\ref{thm:saw4-log} and \ref{thm:WSAWlr} for the WSAW require relatively
minor modifications of the proofs for $|\varphi|^4$.

In the remainder of the paper, we flesh out the above strategy
as it is employed in our context.
To focus on the main ideas, we consider only the
long-range $\varphi^4$ model with $n=1$ component in dimensions $d=1,2,3$.
The essential problem is one of a certain Gaussian integration.

\subsection{Multi-scale Gaussian integration}

Let $d=1,2,3$.
Let $\Lambda$ be the discrete $d$-dimensional torus of period $L^N$,
where $L>1$ is a fixed integer.
The infinite-volume limit is achieved by $N \to \infty$.
Let $C$ be a positive-definite $|\Lambda| \times |\Lambda|$ matrix.
The Gaussian expectation $\Ex_C$ with covariance $C$ of a function
$F:\R^{|\Lambda|} \to \R$ is defined by
\begin{equation}
    \Ex_C F
    =
    \frac{
    \int_{{\mathbb R}^{|\Lambda|}} F(\zeta) e^{-\frac 12 (\zeta,C^{-1}\zeta)}
    \prod_{x\in\Lambda} d\zeta_x
    }{{\int_{{\mathbb R}^{|\Lambda|}}   e^{-\frac 12 (\zeta,C^{-1}\zeta)}
    \prod_{x\in\Lambda} d\zeta_x}}.
\end{equation}

Fix $\alpha \in (0,2)$.
Let $m^2>0$ and let $C$ be the
positive-definite $|\Lambda| \times |\Lambda|$ matrix
\begin{equation}
\label{e:Calpha}
    C = ((-\Delta_\Lambda)^{\alpha/2} + m^2)^{-1}.
\end{equation}
For $g_0>0$ and $\nu_0 \in \mathbb{R}$, let
\begin{equation}
    Z_0=e^{-V_0(\Lambda)},
    \quad
    V_0(\Lambda)
    =
    V_0(\Lambda,\varphi)
    =
    \sum_{x\in\Lambda}
    (\tfrac{1}{4}g_0\varphi_x^4 + \tfrac{1}{2}\nu_0 \varphi_x^2 ).
\end{equation}
The essential problem is to compute the convolution of the Gaussian
expectation $\Ex_C$ with $Z_0$, namely
\begin{equation}
\label{e:ZNdef}
    Z_N(\varphi)=\Ex_CZ_0(\varphi+\zeta) = \Ex_C e^{-V_0(\Lambda,\varphi+\zeta)},
\end{equation}
uniformly as $m^2 \downarrow 0$ and $N \to \infty$.
For example, it is an exercise in calculus to see that the finite-volume
susceptibility is given by
\begin{equation}
\label{e:chiZN}
    \chi_N(g,\nu_0+m^2) = \frac{1}{m^2} + \frac{1}{m^4}\frac{1}{|\Lambda|}
    \frac{D^2Z_N(0;\1,\1)}{Z_N(0)},
\end{equation}
where the directions $\1$ in the directional derivative are the constant
function $\1_x=1$.

To evaluate \eqref{e:ZNdef}, the Gaussian integration is carried out incrementally, or
\emph{progressively}, with each increment effecting integration over a single
length scale.   For this, we use the elementary
property of Gaussian integration that
if $C=C'+C''$ then
\begin{equation}\label{e:progressive2}
  \Ex_C F(\varphi+\zeta) = \Ex_{C''} \Ex_{C'} F(\varphi+\zeta''+\zeta'),
\end{equation}
where on the right-hand side the inner Gaussian integral integrates
with respect to $\zeta'$ (holding $\varphi+\zeta''$ fixed), and  the outer
Gaussian integral then integrates with respect to $\zeta''$.

\begin{figure}[h]
\begin{center}
  \scalebox{0.75}
{\input{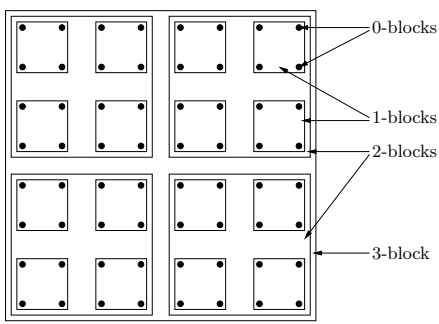}}
\end{center}
\caption{\label{fig:reblock}
Some blocks in ${\cal B}_j$ for
$j=0,1,2,3$, with $d=2$ and $L=2$.}
\label{fig:RG_hierarchy1}
\end{figure}

The choice of $L^N$ as the period of the torus allows for the partition of
the torus into disjoint $j$-\emph{blocks} of side $L^j$, for $j=0,1,\ldots,N$.
The $0$-blocks are simply the points of $\Lambda$, and the unique $N$-block
is $\Lambda$ itself.  In general, the set ${\cal B}_j$ of $j$-blocks
has $L^{(N-j)d}$ elements.
Small values of $j$ are depicted in Figure~\ref{fig:RG_hierarchy1}.
The \emph{scales} $j=0,1,2,\ldots,N$ for the progressive integration
correspond to the block side lengths $L^0,L^1,L^2,\ldots,L^N$.

Given a covariance decomposition
\begin{equation}
    C=((-\Delta_\Lambda)^{\alpha/2} + m^2)^{-1}=\sum_{j=1}^N C_j,
\end{equation}
it follows from \eqref{e:ZNdef} and \eqref{e:progressive2} that
\begin{equation}
\label{e:progressive}
    Z_N(\varphi) =
    \Ex_{C_N+\cdots+C_1}(Z_0(\varphi+\zeta))
    = \Ex_{C_N} \cdots \Ex_{C_2} \Ex_{C_1}(Z_0(\varphi + \zeta_N + \cdots \zeta_1)) .
\end{equation}
We use a carefully constructed covariance decomposition, such that
in the corresponding decomposition $\zeta=\zeta_1+\cdots+\zeta_N$ of the field,
the \emph{fluctuation field} $\zeta_j$ captures the fluctuations of the
field $\zeta$ on scale $j-1$.  This is quantified by estimates on the covariances
in the decomposition,
which express the fact that a typical Gaussian field with covariance
$C_{j+1}$ is roughly constant on $j$-blocks and has size
of order $L^{-j(d-\alpha)/2}$.
These estimates hold until the \emph{mass scale} $j_m$, which is the
smallest value of $j$ for which $L^{\alpha j}m^2 \ge 1$; for scales $j > j_m$
the covariance is smaller and the integrations for such covariances is subject
to a simpler analysis.
An additional finite-range property
of the covariances plays an important
simplifying role by making the field values in
non-contiguous blocks independent
\cite{BGM04,Baue13a,Mitt16,Slad18}.

In view of \eqref{e:progressive}, we define a sequence iteratively by
\begin{equation}
    Z_{j+1}(\varphi)= \Ex_{C_{j+1}}  Z_j (\varphi+\zeta),
    \quad\quad Z_0(\varphi)=e^{-V_0(\Lambda,\varphi)}.
\end{equation}
Each step in the sequence performs integration of a fluctuation field
on a single scale.
Then $Z_N$ is the final element of the sequence, and we are interested in
the limit $N \to \infty$.    We wish to start the
sequence with $Z_0$ defined in terms of $V_0$ with $\nu_0$ slightly above
the critical value $\nu_c$.  However, we do not have a useful
{\it a priori} description of
$\nu_c$; its identification is part of the problem.
To deal with this issue, we enlarge the focus, and consider a Gaussian
convolution as a mapping on a space of functions of the field, defined
on a suitable domain.
In other words, given a covariance $C_+=C_{j+1}$, we write $\Ex_+=\Ex_{C_+}$
and define
a scale-dependent map $Z \mapsto Z_+$ by
\begin{equation}\label{e:ZZ}
  Z_+(\varphi) = \Ex_+ Z(\varphi+\zeta),
\end{equation}
for integrable $Z$.
Given a function $F$ of the field, and given a field $\varphi$,
we define a new function
$\theta_\varphi F$ by $(\theta_\varphi F)(\zeta) = F(\varphi+\zeta)$.
Then we can rewrite \eqref{e:ZZ} compactly as
\begin{equation}\label{e:ZZtheta}
  Z_+ = \Ex_+ \theta Z.
\end{equation}

We wish to capture the scale invariance at the critical point as
a ``fixed point'' of the mapping $Z \mapsto Z_{+}$.  We do not achieve this
literally, because of lattice effects.  Indeed, the mapping is between
\emph{different} spaces, with different norms that implement rescaling.
Nevertheless the notion of a fixed point provides vital guidance.

\subsection{Relevant and irrelevant monomials}

The mapping $Z \mapsto Z_{+}$ is a transformation of one function of the
field to another, and we wish to
identify which are the important aspects of the map to track carefully, and
which parts can be regarded as remainders.

For small $\varphi$, an approximation of $Z(\varphi)$ involves monomials
$\varphi_x^p$.  The relative importance of such monomials is assessed
by calculating their size when summed over a block $B \in {\cal B}_j$, when
$\varphi_x$ is a typical Gaussian field for the covariance $C_+$.
For the specific choice $\alpha = \frac 12 (d+\epsilon)$ in the
covariance \eqref{e:Calpha}, this leads to
\begin{equation}
\label{e:monomials}
    \sum_{x \in B} \varphi_x^p \approx L^{dj}(L^{-j(d-\alpha)/2})^p
    =
    \begin{cases}
        L^{dj} & (p=0)
        \\
        L^{\alpha j} & (p=2)
        \\
        L^{\epsilon j} & (p=4).
    \end{cases}
\end{equation}
For powers $p>4$, a negative power of $L^j$ instead occurs, so such monomials
scale down as the scale is advanced.  The monomials $1,\varphi^2,\varphi^4$
are said to be \emph{relevant} or \emph{expanding},
while $\varphi^6, \varphi^8, \ldots$
are \emph{irrelevant}.  The relevant monomials $\varphi^2,\varphi^4$ appear
already in $V_0$.  The monomial $1$ plays a relatively insignificant
role for the analysis of the susceptibility.
Monomials containing
spatial gradients need also to be considered, in general, but for the long-range
model such monomials are irrelevant.

\subsection{Perturbation theory}

With the classification of monomials as relevant or
irrelevant in mind, we treat $Z$ as approximately equal to
$e^{-V(\Lambda)}$ with $V$ given by a local
polynomial $V(\Lambda) = \sum_{x\in \Lambda}(\frac 14 g\varphi_x^4 +
\frac 12 \nu \varphi_x^2 + u)$
with \emph{coupling constants} $g,\nu,u$.
We seek to find $V_+$, defined with new coupling constants
$g_+,\nu_+,u_+$, such that $Z_+$ is well approximated by $e^{-V_+(\Lambda)}$.
Then the map $Z \mapsto Z_+$ is approximately captured by the map $V \mapsto V_+$.
We refer to $V$ as the \emph{perturbative coordinate}.
The term ``perturbation theory'' refers to the evaluation of the map
$V \mapsto V_+$ to some specific order in $V$, together with
the analysis of this approximate map to compute critical exponents.  We
consider second-order perturbation theory here.

It is straightforward to compute $\Ex_+e^{-\theta V(\Lambda)}$ as a formal power
series in $V$ to within an error of order $V^3$.
Details of a way to do this
are laid out in \cite{BBS-rg-pt}.  Up to irrelevant
terms, the upshot is that
$\Ex_+ e^{-\theta V(\Lambda)} \approx e^{-V_{+}(\Lambda)}$ with
$V_+(\Lambda)=\sum_{x\in\Lambda}(g_+\varphi_x^4+\nu_+\varphi_x^2+u_+)$
and with the  coupling constants $g_+,\nu_+,u_+$
given by an explicit quadratic polynomial in $g,\nu,u$ with
coefficients determined by the covariance $C_+$.

In order to maintain the approximation of $Z$ by $e^{-V}$ over all scales,
a critical
$m^2$-dependent choice $\nu_0=\nu_0^c(m^2)$ is required.
With the wrong choice, $V$ would grow
exponentially and not remain small as the scale advances.
As $m^2\downarrow 0$, $\nu_0(m^2)$ approaches the critical value $\nu_c$.
If we are able to control the above approximations over all scales,
then we finally arrive at $Z_N(\varphi) \approx e^{-V_N(\Lambda,\varphi)}$.
Substitution of this approximation into the right-hand side of \eqref{e:chiZN}
leads, after a small calculation, to
\begin{equation}\label{e:chinu}
  \chi_N(\nu_0^c(m^2)+m^2) \approx \frac{1}{m^2} - \frac{\nu_N(m^2)}{m^4}.
\end{equation}
The proof of Theorem~\ref{thm:phi4lr} in \cite{Slad18}
verifies that the approximation
\eqref{e:chinu} is indeed valid, and that, moreover,
for small $m^2>0$ the following limits hold for general $n$:
\begin{equation}\label{e:nuinfty}
  \lim_{N\to\infty}\nu_N(m^2) = O(m^2\epsilon),
  \quad
   \lim_{N\to\infty} \frac{\partial \nu_N(m^2)}{\partial \nu_0} \Big|_{\nu_0=\nu_0^c(m^2)}
   \asymp m^{2\frac{n+2}{n+8}\frac{\epsilon}{\alpha} + O(\epsilon^2)}.
\end{equation}
Together, \eqref{e:chinu}--\eqref{e:nuinfty} imply the differential inequalities
\begin{equation}\label{e:diffineq}
  \frac{\partial \chi}{d\nu}
  \asymp
  - \chi^{2-\frac{n+2}{n+8}\frac{\epsilon}{\alpha} + O(\epsilon^2)},
\end{equation}
and integration then yields the statement of Theorem~\ref{thm:phi4lr}.

Thus the proof of Theorem~\ref{thm:phi4lr} reduces to the validation
of the approximation \eqref{e:chinu} with careful choice of $\nu_0^c(m^2)$,
and the computation of the limits in \eqref{e:nuinfty}.
The first of these two problems is significantly more difficult than the
second.

A change of variables is helpful to understand the flow
of coupling constants under the RG map.
To incorporate the effect of the growth of relevant monomials
in \eqref{e:monomials}, it is natural
to rescale the coupling constants at scale $j$ as $\hat{g}_j=L^{\epsilon j}g_j$ and
$\hat{\nu}_j=L^{\alpha j}\nu_j$.
A further explicit change of variables $(\hat g, \hat \nu) \mapsto (s,\mu)$
creates a simpler triangular system.
In terms of the new variables, the map $V\mapsto V_+$ is described by
\begin{align}
\label{e:sflow}
    s_{+}& =  L^\epsilon s(1-\beta s)  ,
\\
\label{e:muflow}
    \mu_{+}
    &  =
    L^\alpha \left( 1-{ \frac{n+2}{n+8}}\beta s\right) \mu + \cdots,
\end{align}
with a remainder that does not play an important role.  The coefficient $\beta$ is
given in terms of the accumulated covariance $w_{k} = \sum_{i=1}^k C_i$ by
\begin{equation}
    \beta =\beta_j (m^2)
    =
    (n+8)L^{-\epsilon j} \sum_{x \in \Lambda}(w_{j+1;0,x}^2-w_{j;0,x}^2).
\end{equation}
Properties of the covariance decomposition imply that, for $m^2=0$, the limit
$a=\lim_{j\to\infty}\beta_j(0)$ exists; this permits $\beta$ to be replaced by $a$
in \eqref{e:sflow}--\eqref{e:muflow} up to a controlled error.

The equation $s =  L^\epsilon s(1-a s)$ has two fixed points: an unstable
fixed point $s=0$ and a stable fixed point $\bar s = \frac{1}{a}(1-L^{-\epsilon})$
which is order $\epsilon$.
The perturbative equations \eqref{e:sflow}--\eqref{e:muflow} can be analysed
to conclude that if $s$ is initially close to $\bar s$
then there is an initial choice of $\mu$, which determines $\nu_0^c(m^2)$, from
which equations \eqref{e:sflow}--\eqref{e:muflow} can be iterated indefinitely
and \eqref{e:nuinfty} holds.  The requirement that $s$ be chosen close
to $\bar s$ is a requirement to be initially near the stable fixed point,
and is responsible for the restriction on $g$ in
Theorems~\ref{thm:WSAWlr} and \ref{thm:phi4lr}.

The above analysis is based on the supposition that the approximation
$\Ex_+ e^{-V(\Lambda,\varphi +\zeta)} \approx e^{-V_{+}(\Lambda,\varphi)}$ remains
valid over all scales and over the entire volume $\Lambda$.
This approximation has uncontrolled nonperturbative
errors as
the volume parameter $N$ goes to infinity or as the field $\varphi$ becomes
large.

\subsection{Nonperturbative RG coordinate}

The perturbative coordinate $V$ is supplemented by
a nonperturbative coordinate $K$ which controls all errors in the
above approximations.   A description of $K$ requires the
introduction of the following concepts.

We fix a scale $j$ which we drop from the notation; scale $j+1$ is denoted by
$+$.
A \emph{polymer} is a union (possibly empty) of blocks from ${\cal B}$.  We write ${\cal P}$
for the set of polymers, and ${\cal B}(X)$ and ${\cal P}(X)$ for the
sets of blocks and polymers contained in the polymer $X\in {\cal P}$.
Let ${\cal N}$ denote the algebra of smooth
functions of the field $\varphi$.  We consider maps $F:{\cal P}\to{\cal N}$,
e.g., $F(X)=e^{-V(X)}$ with $V(X,\varphi) =\sum_{x\in X}V(\varphi_x)$.
Given $F, G: {\cal P} \to {\cal N}$,
we define the \emph{circle product} $F \circ G: {\cal P} \to {\cal N}$ by
\begin{equation}
\label{e:FcircG}
  (F \circ G)(X) = \sum_{Y \in {\cal P}(X)} F(Y)G(X\setminus Y)
  \quad\quad (X \in {\cal P}).
\end{equation}
The circle product depends on the scale, since ${\cal P}$
does.  It is commutative and
associative, with unit $\1$ which takes the value $1$ on the empty polymer
and the value $0$ on any nonempty polymer.
We say that $F: {\cal P} \to {\cal N}$
\emph{factorizes over blocks} if $F(X) = \prod_{B\in {\cal B}(X)} F(B)$ for all
$X \in {\cal P}$, e.g., $F(X)=e^{-V(X)}$ factorizes over blocks.
If $F$ and $G$ both factorize over blocks then
  \begin{equation}
  \label{e:binom}
    (F \circ G)(X)
    = \prod_{B\in{\cal B}(X)} (F(B)+G(B))
    \quad\quad (X \in {\cal P}),
  \end{equation}
since in this case expansion of the product on the right-hand side produces
the sum in
\eqref{e:FcircG}.

Instead of the approximation $Z(\Lambda) \approx e^{-V(\Lambda)}$ used in perturbation theory,
we use an exact formula
\begin{equation}
\label{e:ZIcircK}
    Z(\Lambda) = e^{-u|\Lambda|}(I\circ K)(\Lambda).
\end{equation}
Here $I=I(V)$ factorizes over blocks;
it may be regarded for present purposes
as $I(X) = e^{-V(X)}$ but in fact an additional term must be included.
The $K$ appearing on the right-hand side of \eqref{e:ZIcircK}
is
a nonperturbative quantity which encapsulates all errors in perturbation
theory, much as the Taylor remainder formula expresses the error in
a Taylor approximation.
Initially, at scale $0$ we have $Z_0(\Lambda) = e^{-V_0(\Lambda)}
= (e^{-V_0}\circ \1)(\Lambda)$, so \eqref{e:ZIcircK} holds with $K_0=\1$.
We seek to preserve the form
of $Z$ after the Gaussian expectation:
\begin{equation}
\label{e:ZZ+}
    Z_+(\Lambda)
    = \Ex_+\theta Z(\Lambda)
    = e^{-u_+|\Lambda|}(I_+\circ K_+)(\Lambda)
    ,
\end{equation}
with a scale-$(j+1)$ circle product, and with
the operator $\theta$ as in \eqref{e:ZZtheta}.
The choice of $I_+$ is determined by perturbation theory.
Given \emph{any} choice of $I_+$, there is a $K_+$ such that
\eqref{e:ZZ+} holds.  In fact, there are many, as the representation
$Z_+(\Lambda)= e^{-u_+|\Lambda|}(I_+\circ K_+)(\Lambda)$ does not uniquely
determine $K_+$.
The following proposition is a prototype for an effective choice of $K_+$.
For its statement,
the \emph{closure} of a polymer $X\in {\cal P}$ is defined to be
the smallest polymer
$\overline X \in {\cal P}_+$ such that $X \subset \overline X$.

\begin{prop}
\label{prop:reblock}
Suppose that $I,I_+$ factorize over blocks $B \in {\cal B}_j$.
For $X \in {\cal P}_j$,  let  $\delta I (X)= \prod_{B \in {\cal B}(X)}
(\theta I (B) - I_+(B))$.
Then
\begin{equation}
\label{e:EIK1}
    \Ex_{+}\theta (I \circ K)(\Lambda)
    = ( I_+ \circ \tilde{K}_+)(\Lambda)
\end{equation}
with
\begin{equation}
\label{e:EIK2}
    \tilde{K}_+ (U)
=
    \sum_{X \in{\cal P}(U)}
     I_+(U\setminus X)
    \Ex_{+} \big(
    (\delta I
    \circ
    \theta K
    )(X) \big)
    \1_{\overline X =U}
    \quad\quad(U \in {\cal P}_{+}).
\end{equation}
\end{prop}

\begin{proof}
By hypothesis, and by \eqref{e:binom}
at scale $j$ with $F=I_+$ and $G=\delta I$,
\begin{equation}
\label{e:chvar1}
    \theta(I\circ K)
    = (I_+ \circ \delta I) \circ \theta K
    =
    I_+ \circ (\delta I \circ \theta K).
\end{equation}
Let $J = \delta I \circ \theta K$.
Since $I_+$ does not depend on the integration variable (which is introduced only by the
operation $\theta$),
\begin{align}
    \Ex_{+}\theta
    (
    I \circ K  )(\Lambda)
    &   =
       (
    I_+   \circ \Ex_{+} J )(\Lambda)
    =
    \sum_{X \in {\cal P}}
     I_+(\Lambda \setminus X)
    \Ex_{+}\big(J(X) \big)
    .
\end{align}
We reorganize the sum over $X$ by first summing over polymers $U \in {\cal P}_+$
and then summing over all $X\in {\cal P}$ with closure $\overline X = U$.  This gives
\begin{align}
    \Ex_{+}\theta
    (
     I \circ K )(\Lambda)
    & =
    \sum_{U \in {\cal P}_{+}}
     I_+(\Lambda \setminus U)
    \sum_{X \in {\cal P}(U) }
     I_+(U \setminus X)
    \Ex_{+}\big(J(X) \big)
    \1_{\overline X =U}
    .
\end{align}
The right-hand side is \eqref{e:EIK1} with $\tilde{K}_+$ given by \eqref{e:EIK2}, and
the proof is complete.
\end{proof}

The nonperturbative
coordinate $K$ must have two features:  it must be $O(V^3)$,
and it must contract as the scale advances.  Each of these demands requires
a norm on $K:{\cal P}\to {\cal N}$; we do not describe the delicate
choice of norm here \cite{BS-rg-step}.
The $\tilde K_+$ produced by Proposition~\ref{prop:reblock} is a start, but it is
insufficient as it can be shown to be $O(V^2)$ rather than $O(V^3)$,
and neither is it contractive.
Delicate adjustments are required to achieve these two
goals \cite{BS-rg-step}.

On the other hand, $\tilde K_+$ does preserve a good
factorization property.
We say that polymers
$X,Y\in {\cal P}_j$ are \emph{disconnected} if they are separated
by distance at least $L^j$.
A polymer is \emph{connected} if it is
not the union of two disconnected polymers,
and any polymer $X$ partitions into connected components
${\rm Comp}(X)$ which are separated by distance at least $L^j$.
We say that $F: {\cal P} \to {\cal N}$
\emph{factorizes over connected components}
if $F(X) = \prod_{Y \in {\rm Comp}(X)} F(Y)$.
The finite-range property of the covariance decomposition is the statement
that $C_{j;xy}=0$ if $\|x-y\|_1 \ge \frac 12 L^j$.
This ensures that
\begin{equation} \label{e:factor}
  \Ex_{+}\big(F(X)G(Y)\big) = \Ex_{+}\big(F(X)\big)\Ex_{+}\big(G(Y)\big)
  \quad
  \text{if $X,Y \in {\cal P}_+$ are disconnected,}
\end{equation}
because uncorrelated Gaussian random variables are independent.

\begin{figure}[h]
\begin{center}
\includegraphics[scale = 0.18]{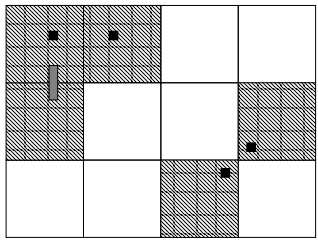}
\end{center}
\caption{\label{reblock-euc}
The five large shaded blocks represent $U$, which is the closure of the
polymer $X$ consisting of the four small dark blocks (the support of $\delta I$) and the
small shaded rectangle (the support of $K$).}
\end{figure}

Suppose that $K$ factorizes over connected components at scale $j$.
It can be verified that $\tilde{K}_+$ then factorizes over connected components
at scale $j+1$, using \eqref{e:factor}.  This factorization
functions in parallel with the norm, which
has the property that the norm of a product is at most the product of the norms.
The geometry of the identity \eqref{e:EIK2} defining $\tilde{K}_+(U)$ is illustrated in Figure~\ref{reblock-euc}, which is helpful for the verification of factorization.

\subsection{RG map and phase portrait}

The RG map is a scale-dependent map
\begin{equation}
\label{e:RGmap}
    {\rm RG} :(s,\mu  , K)
    \mapsto
    (s_{+} , \mu_{+}  , K_{+}),
\end{equation}
defined on a suitable domain.  It is defined in such a way that $K_{+}$ is
third order in $(s,\mu)$ if $K$ is, and the $K$ component is
contractive under change of scale.
The values of $(s_{+},\mu_{+})$ depend on $K$ as well as
$(s,\mu)$, and this dependence is engineered to remove the
relevant parts from $K$.  This extraction is responsible
for the contraction of $K$ under change of scale, and is
indispensable
for the iteration of the RG map over all scales.  As long as $K$ is
third order, its effect on the flow of the coupling
constants does not change the second order perturbation theory that determines
the asymptotic
behaviour of $\nu_N$ and the critical exponent $\gamma$ for the susceptibility.

\begin{figure}[h]
\begin{center}
\includegraphics[scale = 0.405]{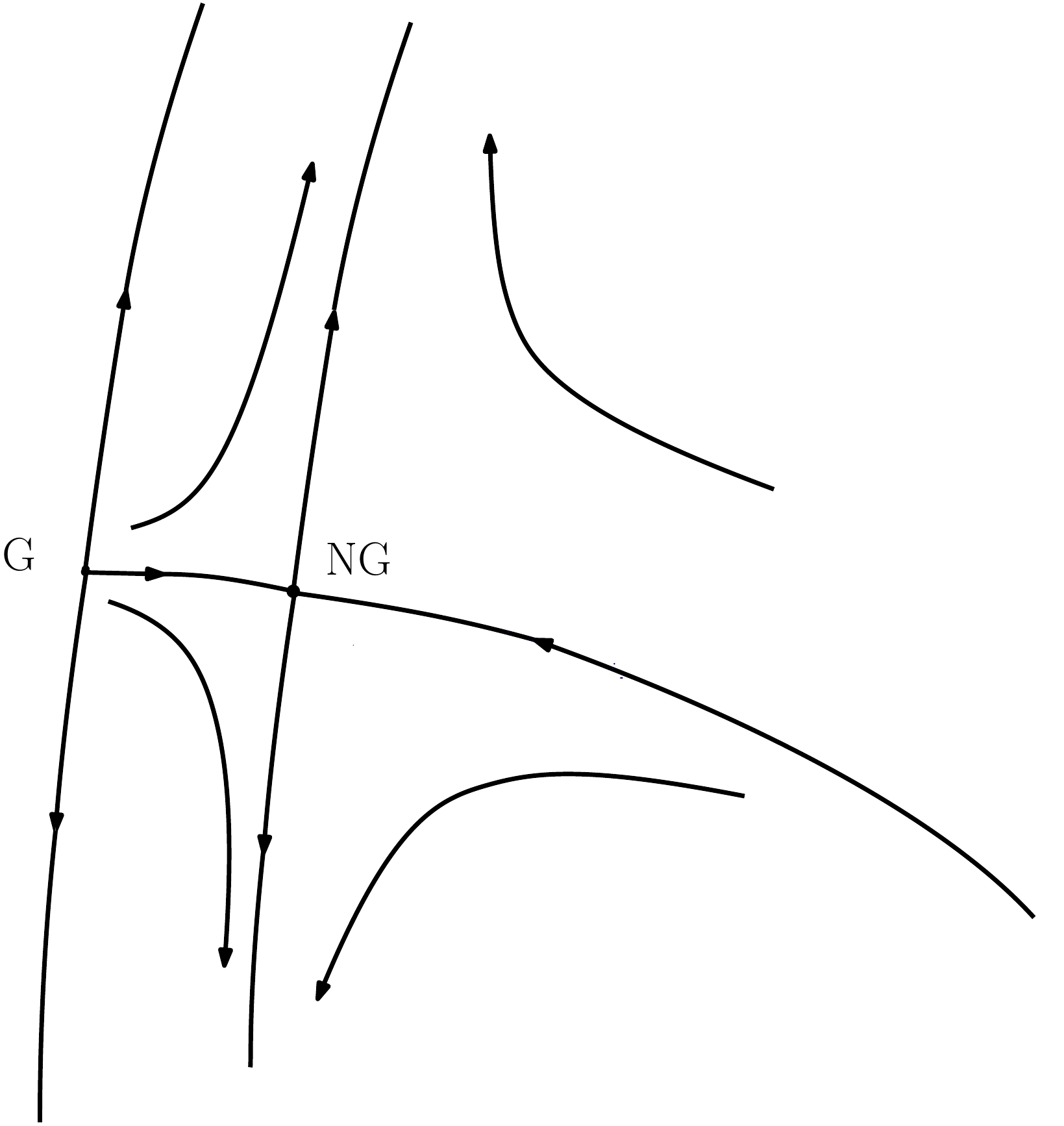}
\end{center}
\caption{\label{fig:phase}
Schematic phase portrait of the dynamical system.}
\end{figure}

The RG map is used to define a map $T$
on a space of sequences $(s_j,\mu_j,K_j)_{j \ge 0}$, such that a fixed point of
$T$ corresponds to a sequence which provides a recursive solution to
\eqref{e:RGmap} for all scales $j$.  The $j=0$ value
of this \emph{global RG flow} identifies the
critical point $\nu_c$.  The RG flow is depicted schematically
by the phase portrait shown in Figure~\ref{fig:phase}.
For the long-range model with $\alpha=\frac 12(d+\epsilon)$ in dimensions $d=1,2,3$,
the non-Gaussian Wilson--Fisher hyperbolic
fixed point is stable and the Gaussian fixed
point is unstable.  The critical point lies on the stable manifold from
which the flow converges to the non-Gaussian fixed point.  The 1-dimensional
unstable manifold reflects the growth of $\mu$ for a non-critical choice of
initial condition.
For the 4-dimensional nearest-neighbour model, the two fixed points merge
into a single stable non-hyperbolic Gaussian fixed point.

\section{Conclusion}

The creation of a comprehensive
theory of phase transitions and critical phenomena is one of
the great achievements of theoretical physics during the second half of the
last century.
The mathematical problems posed by that theory remain a very active topic of
current research.  Wilson's RG approach
is a cornerstone of the physical theory.  Mathematical theorems based on
the RG approach began to appear decades ago, but a great deal remains to
be done to provide a complete and  nonperturbative understanding
of critical phenomena, without
uncontrolled approximations.
This paper concerns some recent contributions in this direction,
for the weakly self-avoiding walk and the $|\varphi|^4$ lattice spin model,
including a rigorous version of the $\epsilon$-expansion.

\section*{Acknowledgements}

I am grateful to David Brydges and Roland Bauerschmidt for all I
have learned from them
during our long collaborations; their influence
runs throughout the paper.
I thank both of them for helpful suggestions,
and Nathan Clisby for providing
Figures~\ref{fig:saws}--\ref{fig:long-range-srw}.
Figures~\ref{fig:ising-arrows}--\ref{reblock-euc}
are taken from \cite{BBS-brief}.

\end{document}